\documentclass[fleqn,usenatbib,aas_macros]{mnras}

\usepackage{newtxtext,newtxmath}

\usepackage[T1]{fontenc}
\usepackage{ae,aecompl}
\usepackage{tikz}
\usetikzlibrary{calc}

\usepackage{graphicx}	
\usepackage{amsmath}	
\usepackage{amssymb}	
\usepackage{float}
\usepackage{hyperref}

\hypersetup{draft}
\hypersetup{
colorlinks,
citecolor=black,
filecolor=black,
linkcolor=black,
urlcolor=black
}

\newcommand*{\unsim}{\mathord{\sim}} 

\begin{document}


\title[PTF1 J085713+331843]{PTF1 J085713+331843, a new post common-envelope binary in the orbital period gap of cataclysmic variables}
\author[J. van Roestel]
{J. van Roestel$^1$\thanks{j.vanroestel@astro.ru.nl},
P.J. Groot$^1$,
D. Levitan$^2$,
T.A. Prince$^2$,
S. Bloemen$^1$,
\newauthor 
T.R. Marsh$^3$,
V.S. Dhillon$^{4, 5}$,
D. Shupe$^6$, 
R. Laher$^7$\\
$^1$Department of Astrophysics/IMAPP, Radboud University Nijmegen, P.O.Box 9010, 6500 GL, Nijmegen, NL \\
$^2$Cahill  Center  for  Astronomy  and  Astrophysics,  California
Institute of Technology, Pasadena, CA 91125, USA \\
$^3$Department of Physics, University of Warwick, Gibbet Hill Road, Coventry, CV4 7AL, UK \\
$^4$Department of Physics and Astronomy, University of Sheffield, Sheffield S3 7RH, UK \\
$^5$Instituto de Astrof\'isica de Canarias, E-38205 La Laguna, Tenerife, Spain \\
$^6$Infrared Processing and Analysis Center, California Institute
of Technology, Pasadena, CA 91125, USA \\
$^7$Spitzer Science Center, California Institute
of Technology, Pasadena, CA 91125, USA}
\date{Accepted Year Month Day. Received Year Month Day; in original form Year Month Day}

\pagerange{\pageref{firstpage}--\pageref{lastpage}} \pubyear{Year}

\maketitle

\label{firstpage}

\maketitle
\begin{abstract}
We report the discovery and analysis of PTF1 J085713+331843, a new eclipsing post common-envelope detached white-dwarf red-dwarf binary with a 2.5h orbital period discovered by the Palomar Transient Factory. ULTRACAM multicolour photometry over multiple orbital periods reveals a light curve with a deep flat-bottomed primary eclipse and a strong reflection effect. Phase-resolved spectroscopy shows broad Balmer absorption lines from the DA white dwarf and phase-dependent Balmer emission lines originating on the irradiated side of the red dwarf. The temperature of the DA white dwarf is $T_\mathrm{WD} = 25700 \pm 400\,$K and the spectral type of the red dwarf is M3-5. A combined modelling of the light curve and the radial velocity variations results in a white dwarf mass of $M_\mathrm{WD} = 0.61^{+0.18}_{-0.17}\, \mathrm{M_{\odot}}$ and radius of $R_\mathrm{WD} = 0.0175^{+0.0012}_{-0.0011}\, \mathrm{R_{\odot}}$, and a red dwarf mass and radius of $M_\mathrm{RD} = 0.19^{+0.10}_{-0.08}\, \mathrm{M_{\odot}}$ and $R_\mathrm{RD} = 0.24^{+0.04}_{-0.04}\, \mathrm{R_{\odot}}$. The system is either a detached cataclysmic variable or has emerged like from the common envelope phase at nearly its current orbital period. In $\unsim70\,$Myr, this system will become a cataclysmic variable in the period gap.
\end{abstract}

\begin{keywords}
binaries: eclipsing -- stars: white dwarfs -- late-type
\end{keywords}

\section{Introduction}
The majority of stars are members of binary systems. In a main sequence binary with an initial separation less than $\unsim1000$ solar radii, the two components of the system will interact during their evolution.  When the more massive star ascends the red giant branch and/or asymptotic giant branch, it engulfs the secondary star in a common envelope (\citealt{1976IAUS...73...75P}; for reviews see \citealt{2000ARA&A..38..113T,2008ASSL..352..233W,2010NewAR..54...65T,2013A&ARv..21...59I}). During this phase the system loses orbital angular momentum, causing the secondary star to spiral inward and the giant's envelope to be expelled. This process is expected to take up to a few hundred years at most, resulting in a binary with a short orbital period, consisting of the core of the primary star (now a white dwarf or subdwarf B/O star) and the main-sequence secondary star. These systems are known as post-common-envelope binaries (PCEBs). 

Eclipsing PCEBs are ideal to measure fundamental system parameters, such as the mass and radius of both components, with a high accuracy and independent of stellar atmosphere models. In addition, the sharp eclipses allow for very accurate orbital period measurements. Only 71 eclipsing white dwarf PCEBs systems are known \citep{2015MNRAS.449.2194P}, and, in addition, 14 eclipsing subdwarf B (sdB) binaries were presented in \citet{2015A&A...576A..44K}. NN Serpentis is one of the brightest eclipsing PCEBs and has been studied in most detail. \citet{Parsons:2009} determined the masses and radii of both components with an uncertainty of $\leq$4\%. Eclipse timing studies of NN Ser also revealed periodic deviations of the expected eclipse times, which can be explained by two circumbinary planets \citep{2013A&A...555A.133B,2014MNRAS.437..475M}. Other detailed studies of individual eclipsing PCEBs have been presented by \citet{Pyrzas:2011bk,Parsons:2011a,2012MNRAS.420.3281P,2012MNRAS.426.1950P,2015ApJ...808..179D}. Besides being interesting individually, the population of PCEBs puts constraints on the evolutionary stages of detached white-dwarf red-dwarf binary systems.

After the common-envelope phase, the system will subsequently lose orbital angular momentum through magnetic braking and/or gravitational wave radiation. If the red dwarf secondary fills its Roche lobe while it is still a main-sequence star, a cataclysmic variable (CV) is formed (see \citealt{2001ApJ...550..897H,Knigge:2011}, for a detailed analysis of CV evolution). A statistically significant lack of CV systems has been observed in the period range between $\approx$2.15 and $\approx$3.18h; the so-called period gap \citep{2003A&A...404..301R,2009MNRAS.397.2170G}. The disrupted magnetic braking model \citep{1983ApJ...275..713R,2010A&A...513L...7S} predicts that mass transfer stops at an orbital period of $\approx$3.18h and resumes again at an orbital period of $\approx$2.15h (see \citet{Knigge:2011}). Since the passage through the period gap is driven by gravitational wave radiation only, the space density of systems in the gap should be higher than that of systems just above and below the gap if all CVs start mass transfer above the gap and then evolve through the orbital period gap. The relative number of hibernating (in-gap) cataclysmic variables to regular PCEBs is uncertain. \citet{2008MNRAS.389.1563D} predict that detached CVs outnumber the regular white dwarf PCEBs by a ratio of 4 to 13, with a pile up at the high end of the period gap since gravitational wave emission strength is a strong function of orbital period \citep[$\dot{P}\propto P^{-5/3}$, see][]{1963PhRv..131..435P,
1964PhRv..136.1224P}. A recent study by \citet{2016arXiv160107785Z} finds PCEBs with orbital periods between $\approx$2.15 and $\approx$3.18h at a rate higher than which can be explained with standard PCEB formation theories, requiring a fraction of systems to be hibernating CVs.

In this paper we report on PTF1 J085713+331843 (PTF0857), a new PCEB discovered by the Palomar Transient Factory (PTF). PTF0857 was selected as a cataclysmic variable candidate because of its colours in the Sloan Digital Sky Survey (SDSS, \citealt{2012ApJS..203...21A}). A visual inspection of the PTF light curve showed that, on a period of a few hours, PTF0857 decreased in brightness by more than 1 magnitude. This was confirmed by inspecting the PTF images and the system was targeted for follow-up with the Double Beam Spectrograph at the Palomar $200^{\prime\prime}$ (P200) and the triple beam, high cadence imager ULTRACAM at the 4.2m William Herschel Telescope (WHT) (Section~\ref{sec:observations}). Our analysis of the data is explained in Section~\ref{sec:analysis}. Section~\ref{sec:discussion} presents the results and the determination of the system parameters. In Section~\ref{sec:evolution} we discuss the past and future evolution of PTF0857.

\section{Observations}\label{sec:observations}
\begin{figure}
\begin{center}
\includegraphics[width=84mm]{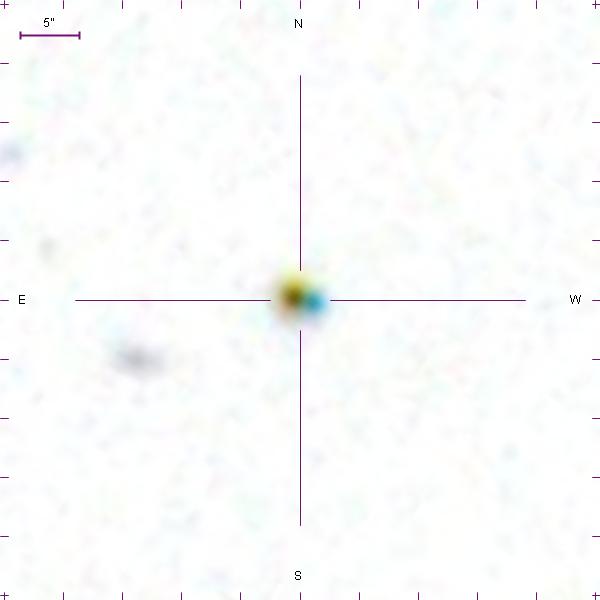}
\caption{SDSS image of PTF0857 with the colours inverted. By eye two objects can clearly be distinguished, the blue object in the east is PTF0857 and the red object in the west is the interloper. The angular distance between the two objects is $1.46\pm0.10\,$arcsec.}
\label{fig:SDSS}
\end{center}
\end{figure}

\begin{figure*}
\begin{center}
\includegraphics[width=\textwidth]{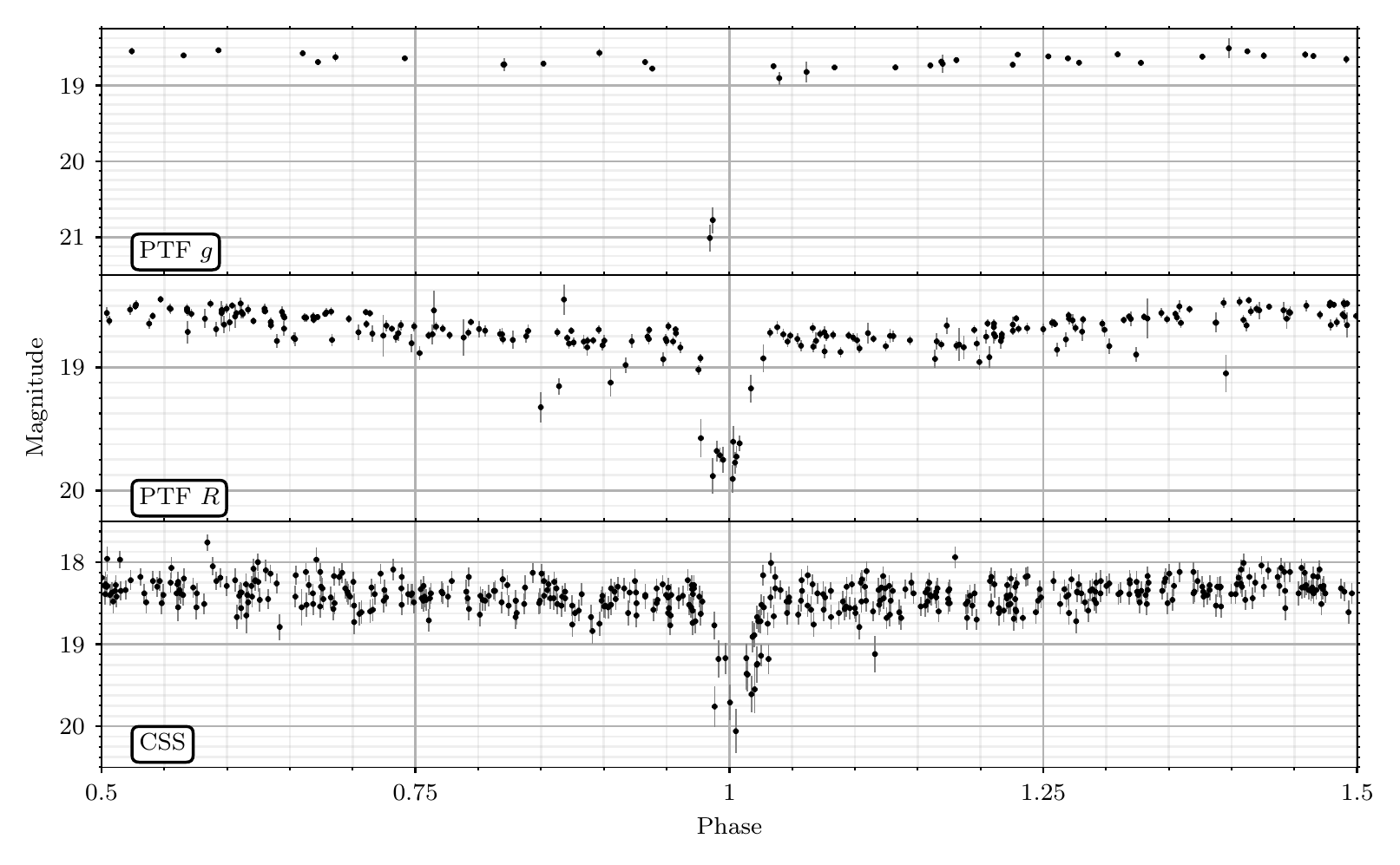}
\caption{Light curves of PTF0857 in PTF R, PTF g, and CSS, folded to the orbital period, 0.10602727 days. The primary eclipse can be clearly seen in all three light curves, as well the small variation due to the reflection effect, best seen in PTF R. }
\label{fig:PTF_CSS_lc}
\end{center}
\end{figure*}

\subsection{SDSS photometry}
A close inspection of the field of PTF0857 shows that the target is blended with another star, referred to as `interloper' in the rest of the paper, see Fig.~\ref{fig:SDSS}. This caused the SDSS pipeline to mis-classify the target as a galaxy. The colours in the SDSS catalogue are thus a combined colour of the two objects. The stars are separated by 1.5 arcsec (see Sect.\ \ref{subsec:mags}), sufficient to determine their individual magnitudes by fitting the PSF of both stars.

\subsection{Palomar Transient Factory photometry}
The PTF \citep{Law:2009,Rau:2009} uses the 48-inch (1.2m) Samuel Oschin Telescope at Palomar Observatory to survey the sky in search of transients. The pixel scale is $1.02\, \mathrm{arcsec\, pixel^{-1}}$ and with a field of view of $7.26\,\mathrm{deg^2}$, it surveys $\approx$2000\,$\mathrm{deg^2}$ per night. Although designed as a transient experiment, the observations are also used to study variable stars. All data are automatically processed to provide light curves, see \citet{2014PASP..126..674L} for further details.

PTF observed the area of the sky containing PTF0857 at a highly irregular cadence and obtained 238 $R_{\textrm{mould}}$-band and 38 $g$-band measurements between 2009 March and 2014 May with an individual exposure time of 60s for all images. Due to a typical seeing of $\unsim2$ arcsec in the PTF images, PTF0857 and the interloper are regarded as a single target. We present the PTF light curves in Fig.~\ref{fig:PTF_CSS_lc}.

\subsection{Catalina Sky Survey photometry}
We examined data obtained by the Catalina Real Time Surveys (CRTS, \citealt{2009ApJ...696..870D,2011arXiv1102.5004D}). One of their telescopes, the Catalina Schmidt Telescope, observed PTF0857 468 times between 2005 April and 2013 May. The telescope is a 0.68 m Schmidt telescope with a pixel scale of 2.5 arcsec\ pixel$^{-1}$. All exposure times are 30\,s. No filter is used to maximize the signal, but the observations are calibrated to $V$-band magnitudes. Observations during a given night are typically grouped in four epochs, 15 minutes apart. The overall cadence varies, but the field is typically observed every $\unsim$20\, d, when visible and weather permitting.

\subsection{ULTRACAM high cadence photometry}
We obtained high-speed photometry of PTF0857 on the 30th and 31st of 2012 January using ULTRACAM \citep{2007MNRAS.378..825D}, mounted on the William Herschel Telescope (WHT) at the Roque de los Muchachos Observatory on the island of La Palma, Spain. ULTRACAM is a triple-beam camera that uses frame-transfer CCDs to minimise dead time. We used 2x2 binning for both nights to reduce the dead-time to 0.024\,s. The first night we obtained 5.5 hours of images in the $u^\prime$, $g^\prime$ and $r^\prime$ filters and the second night 5.2 hours in the $u^\prime$, $g^\prime$ and $i^\prime$ filters. The first night we used exposure times of 3.03\,s in $g^\prime$ and $r^\prime$, and 9.13\,s in $u^\prime$, the second night we used 2.03\,s exposures in $g^\prime$ and $i^\prime$, and 8.18\,s in $u^\prime$. Both nights were photometric, with a stable seeing of $\mathrm{\unsim1.5}$ arcsec, except for the last two\,h of the second night, when the seeing became more unstable and increased to 2.5 arcsec.

All data were reduced using the ULTRACAM pipeline software \citep[see][]{2004MNRAS.355....1F}. Debiasing, flatfielding and background subtraction were performed in the standard way. We used two overlapping, variable size apertures covering both the target and the interloper with a radius of two times the full width at half maximum (FWHM) of the PSF, which itself was determined using two reference stars on the same chip. We calibrated the light curves directly using two stars in the field for which SDSS photometry is available (located at 08:57:30.69 +33:16:47.09 and 08:57:28.82 +33:16:46.08, J2000). The difference in photometric calibration between the two stars in less than 1 percent.

\subsection{Spectroscopy}
Spectra were taken on the nights of 2012 January 30 to 2012 February 1 with the 5.1\,m Hale telescope at Palomar Observatory using the double beam spectrograph DBSP \citep{1982PASP...94..586O}. The seeing was around 1 arcsec and conditions were not photometric. At times clouds prevented observations altogether. A 316 $\mathrm{lines\, mm^{-1}}$ grating, blazed at 7500\AA\ was used in the red arm and a 600 $\mathrm{lines\, mm^{-1}}$ grating, blazed at 4000\AA\ in the blue arm. The wavelength ranges for the blue and the red arms are 3200--5800 and 5200--10500 \AA, respectively, with a resolving power of $R\approx 1400$ in both arms. A total of 41 spectra were obtained in the blue arm, and 36 in the red arm with exposure times of 5\,m. The slit was positioned such that the interloper was not included. In addition, one spectrum in the red arm was taken with the slit only on the interloper.  

The programme \textsc{L.A.Cosmic} \citep{vanDokkum:2001bp} was used to remove cosmic rays, and the spectra were extracted and calibrated using {\sc iraf}. For the wavelength calibration FeAr and HeNeAr lamp spectra taken at the same position of the target were used. For relative flux calibration standard star spectra obtained at the beginning and end of the night were used. The absolute calibration of the spectra was done using the ULTRACAM photometry for each spectrum individually. We convolved the individual spectra with the ULTRACAM response curves and a typical atmospheric absorption curve. We then multiplied the spectra so the total flux matched the ULTRACAM observations at the same orbital phase. For the blue arm we used the $g^\prime$\,band and for the red arm we used the $r^\prime$\,band. The signal-to-noise ratios per \AA ngstr\"om of the spectra vary because of changing conditions, but are typically $\unsim$10 in the blue and $\unsim$7 in the red. 

The short wavelength end of the blue spectra, below $\mathrm{\unsim4200\,}$\AA\, is affected by instrumental defects, which mimic broad absorption lines. Although consistently present in blue DBSP spectra, they are of indeterminate origin. We mask out the affected wavelength ranges in the further analysis.

\section{Analysis}\label{sec:analysis}

\begin{figure}
\begin{center}
\includegraphics[width=84mm]{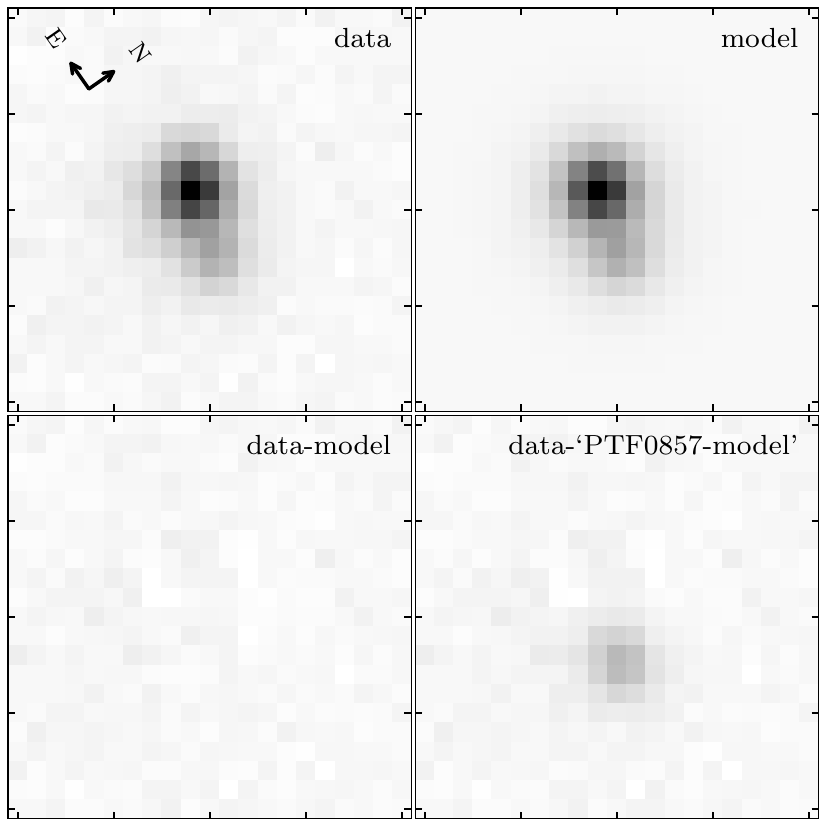}
\caption{The PSF fit to the SDSS $r$ image, with to the east PTF0857 and to the west the interloper. The top left panel shows the data, the top right panel shows the model, the bottom left panel shows the difference between the model and the data, and the bottom right panel shows the difference between the data and the PSF model of PTF0857 only. The greyscale is the same in all images, spanning from minimum to maximum with a linear scale. The size of the images is 21 by 21 pixels, with a pixel scale of $0.39\, \mathrm{arcsec\, pixel^{-1}}$}
\label{fig:PSFfit}
\end{center}
\end{figure}

\subsection{Magnitudes}\label{subsec:mags}
To extract the luminosity of PTF0857 and the interloper from the $ugriz$- SDSS images, we use a custom-written point-spread-function (PSF) fitting code to fit a Moffat profile \citep{1969A&A.....3..455M} to both stars. Both stars have the same parameters, except for position and brightness. We determine the uncertainties on our fits using an implementation of the Affine-Invariant Markov Chain Monte Carlo method from the {\sc emcee} python package \citep{2013PASP..125..306F}. We inspected the residuals of our best models to the images and could not find any pattern in the remaining noise. The results are given in Table~\ref{tab:mags} and Fig.~\ref{fig:PSFfit}. The angular distance between the two stars thus determined is $1.46\pm0.10$\,arcsec.

To get a more accurate measurement of the interloper's magnitudes we fitted the spatial brightness profile in the in-eclipse images using the same procedure as used to determine the SDSS magnitudes, where we fixed the positional offset between PTF0857 and the interloper. The ULTRACAM data have a larger pixel scale compared to SDSS (0.6 arcsec\,pixel$^-1$ versus 0.39 arcsec\,pixel$^-1$), but the advantage is that the interloper is brighter than PTF0857 while in eclipse. When inspecting the residuals we noticed some systematic deviations from zero in the spatial brightness distributions. Because of this, we used bootstrapping \citep[p. 147]{2012psa..book.....W} to determine the uncertainties on the fit, which turned out to be a factor $\unsim3$ higher than the formal uncertainties. We have adopted the uncertainties on the photometry obtained using the bootstrap method. The results are shown in Table~\ref{tab:mags}.

\begin{table*}
  \caption{The coordinates (J2000) and magnitudes (AB) for PTF0857 and the interloper measured from the SDSS images and ULTRACAM images. The numbers in square brackets indicate the magnitude range of PTF0857 due to the reflection effect. The SDSS images were obtained at an unknown phase, so they should fall somewhere in the range of the ULTRACAM magnitudes. The uncertainties do not include any uncertainties in the calibration, which are typically 0.01 mag. The filters used by SDSS are the $ugriz$ system, while ULTRACAM uses the $u^\prime g^\prime r^\prime i^\prime z^\prime$ system. The difference between the systems is small enough that we can neglect it.}
  \centering
 \label{tab:mags}
 \begin{tabular}{lllll}
 & \multicolumn{2}{|c|}{SDSS} & \multicolumn{2}{|c|}{ULTRACAM} \\
 \hline
 & PTF0857&Interloper & PTF0857 & Interloper\\
 \hline
 RA & 08:57:13.274(2) & 08:57:13.162(8)  &  &  \\
 Dec & 33:18:43.11(6) & 33:18:42.74(10)  &  &  \\

  $u/u^\prime$ & $18.59 \pm 0.01$ &  & $[18.53-18.64] \pm 0.01$ &  \\
  $g/g^\prime$ & $18.62 \pm 0.01$ & $22.13 \pm 0.12$ & $[18.66--18.80] \pm 0.01\ $ & $21.18 \pm 0.06$ \\
  $r/r^\prime$ & $18.83 \pm 0.03$ & $20.20 \pm 0.03$ & $[18.85--19.15] \pm 0.02\ $ & $19.98 \pm 0.04$ \\
  $i/i^\prime$ & $18.90 \pm 0.03$ & $19.00 \pm 0.03$ & $[18.84--19.23] \pm 0.04\ $ & $18.90 \pm 0.03$ \\  
  $z/z^\prime$ & $18.92 \pm 0.02$ & $18.41 \pm 0.02$ &  &  \\
  \hline
 \end{tabular}
\end{table*}

\subsection{Orbital period} \label{subsec:ephem}

We determine the orbital period (0.106\,027\,27(4) d) of the binary system using the ULTRACAM, PTF and CRTS data. All our time measurements were converted to MJD at the Solar system barycentre (BMJD), and they are on barycentric dynamical time (TDB), see \citet{2010PASP..122..935E}. For the ULTRACAM light curve, we determine an initial orbital period by fitting a model (Section \ref{subsec:LC}) to the $g^\prime$ band data, resulting in $P_\mathrm{orb, initial}=0.106027(26)$\,d. While this is just an initial measurement, it can be used to break the degeneracy between alias frequencies in the PTF and CRTS data. We use the PTF $g$, PTF $R$, and CRTS light curves to determine the orbital period using a multi-harmonic, multi-band Lomb-Scargle model with parameters $N_\mathrm{base}=0$ and $N_\mathrm{band}=15$ \citep{VanDerPlas:2015}. We bootstrap the data to calculate the uncertainty on the period. The orbital period (assuming it is constant) is $P_\mathrm{orb}=0.10602727(4)$\,d. 

We determine the time of mid-eclipse by modelling the ULTRACAM light curves (see Section \ref{subsec:LC}). The resulting linear ephemeris is given by:
\begin{equation}
\mathrm{BMJD(TDB)} = 55957.121\,914(3) + 0.106\,027\,27(4)\, E \nonumber \\
\end{equation}

All photometry and spectroscopy are folded on this ephemeris. The folded light curves are shown in Fig.~\ref{fig:PTF_CSS_lc}.

\subsection{Spectral type and temperature}\label{subsec:spectemp}
\begin{figure*}
\begin{center}
\includegraphics[width=\textwidth]{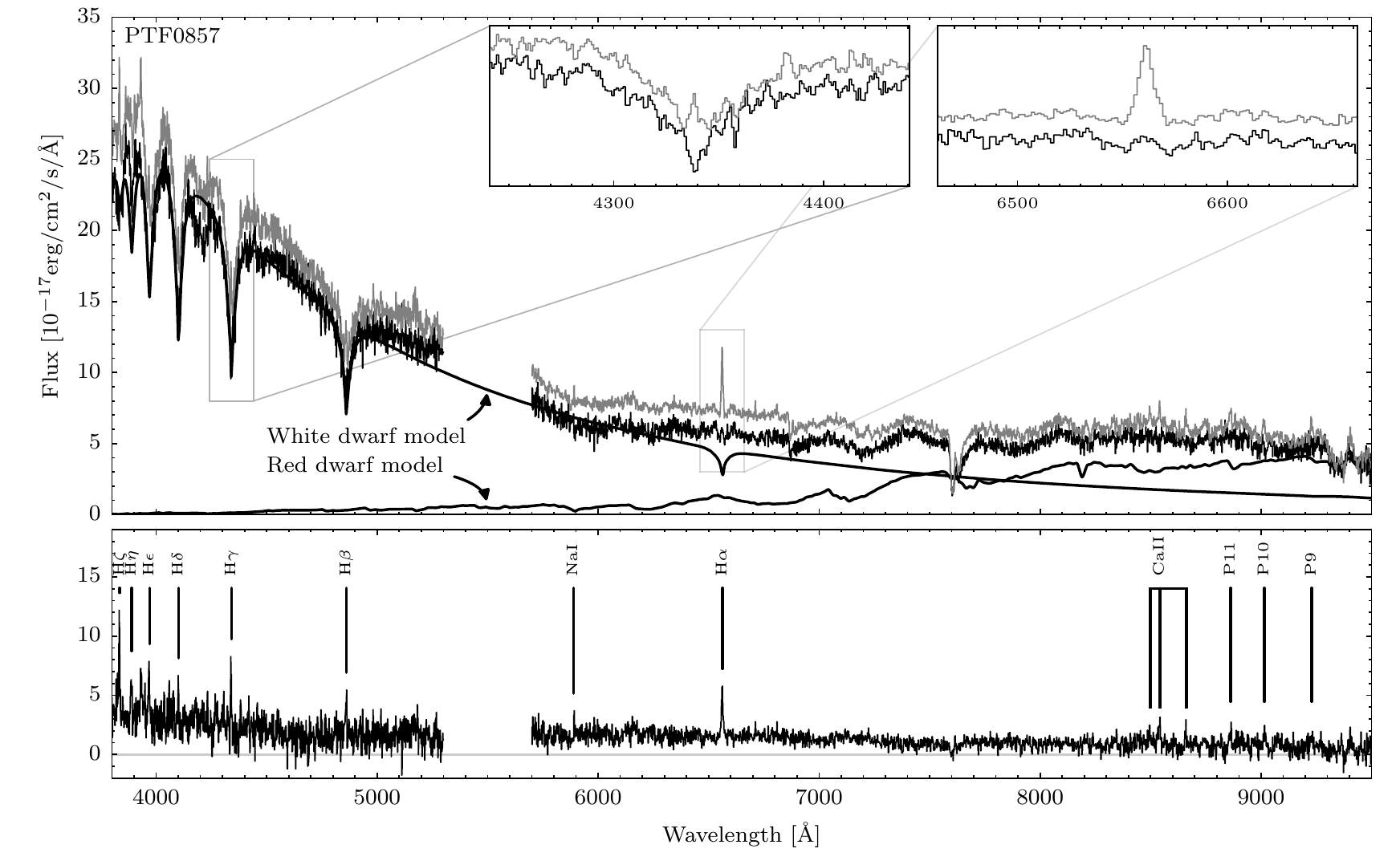}
\caption{Average spectrum of PTF0857 for phases 0.8--0.2 (top,black) and for phases between 0.4 and 0.6 (top,grey). The blue side is dominated by the white dwarf and shows wide Balmer absorption lines. The red part of the spectrum is dominated by the red dwarf as can be seen by the TiO bands and slightly rising flux towards longer wavelengths. To guide the eye, the best fit model spectra are plotted (black lines). The insets show the H$\alpha$ and H$\gamma$ lines in more detail. The bottom panel shows the difference between the spectra taken between phase 0.2-0.8 and 0.4-0.6. It clearly shows phase-dependent Balmer emission lines with H$\alpha$ the most prominent. Weaker lines in the difference spectra are the Paschen lines (9,10,11), the Na-I lines at 5889/5895\AA\ and the Ca-II triplet (8498, 8542 and 8662\AA).}
\label{fig:averagespectrum}
\end{center}
\end{figure*}

The average spectrum is shown in Fig.~\ref{fig:averagespectrum}. The spectral shape is a composite of a blue and red source. The blue spectra are dominated by the white dwarf, as indicated by the broad Balmer absorption lines. The red spectral range is dominated by the red dwarf; the flux increases to higher wavelength and the TiO absorption bands can clearly be seen. The H$\alpha$ absorption line of the white dwarf is barely distinguishable, and a strong, phase-dependent H$\alpha$ emission line is observed (see the inset of Fig. \ref{fig:averagespectrum}). These narrow Balmer emission lines generally originate on the irradiated side of the red dwarf, with the orbital motion causing the variability in observed emission strength as the irradiated hemisphere rotates in and out of the observer's view.

Since no helium absorption lines are seen in the spectrum, we conclude that the white dwarf has a hydrogen dominated atmosphere [DA, for examples of DB spectra, see \citet{2011ApJ...737...28B}]. To determine the white dwarf's temperature and surface gravity and the red dwarf's spectral type, we used the model spectra as in \citet{verbeek:2014}; a combination of DA white dwarf model-spectra from \citet{koester:2001} and observed M-dwarf spectra from \citet{Pickles:1998}, calibrated using fluxes from \citet{Beuermann:2006} (with an uncertainty of 10\% on the luminosities). The white dwarf spectra available to use are in a grid with temperature steps of about 10 percent and $\log{g}$ intervals of 0.5 dex. We use bilinear interpolation (in temperature and surface gravity) to make the white dwarf atmosphere model grid continuous, and we use linear interpolation for the red dwarf spectral types. We do not use any reddening correction, since dust extinction is negligible in the direction of PTF0857 ($E_{B-V}=0.0263$; \citealp{1998ApJ...500..525S}). We use $\mathrm{\chi^2}$ minimization with, as free parameters, the red dwarf spectral type, the white dwarf temperature and its surface gravity, and independent distances for the white dwarf and the red dwarf. To determine the uncertainties on the parameters, we again use {\sc emcee} \citep{2013PASP..125..306F}. The uncertainties reported are only the formal uncertainties from our fit. Uncertainties due to interpolation are not taken into account.

To obtain the white dwarf temperature we fit a white dwarf model to the average of the blue spectra ($\mathrm{\lambda<5500}$ \AA). We only use spectra taken between phases 0.8 and 0.2 (but excluding the eclipse phases from 0.95 to 0.05) because at these phases the contribution by irradiation or reflection is minimal. Although the contribution by the red dwarf in the blue part of the spectrum is expected to be small, we do include a red dwarf component in our model, and treat it as a nuisance parameter. The red dwarf also produces emission lines that could introduce a systematic uncertainty in the result and therefore we mask $\mathrm{80}$\AA\ around each Balmer line. This reduces the accuracy of the fit since we ignore the centre of the white dwarf absorption lines. The models fit the data relatively well ($\mathrm{\chi_{\rm red}^2=1.08}$) with a white dwarf temperature of $\mathrm{25700\pm{400}\,K}$ and surface gravity of $\log g \mathrm{= 7.86^{+0.06}_{-0.07}}$ at a distance of $\mathrm{567^{+19}_{-17}}$\,pc.

\begin{figure}
\begin{center}
\includegraphics[width=84mm]{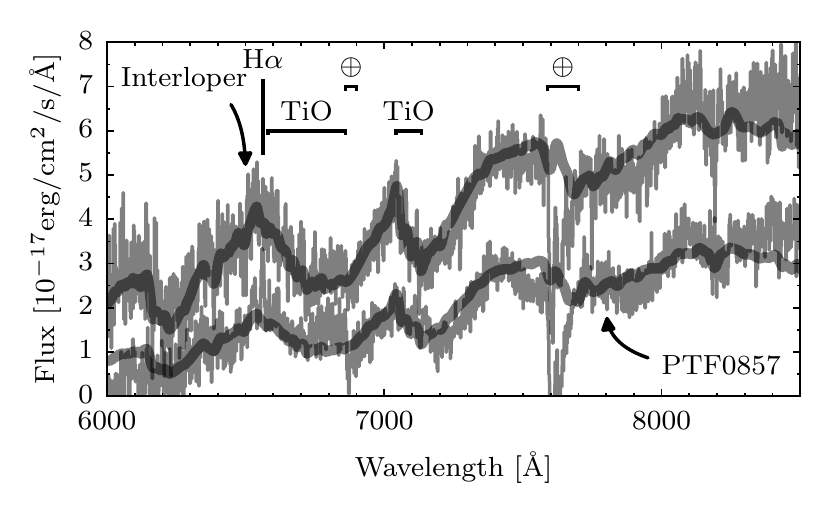}
\caption{The spectrum of the interloper (upper) and the spectrum of PTF0857 after subtracting the best--fitting white dwarf model (lower). The best fitting red dwarf model spectra are over plotted.}
\label{fig:Red_spectrafit}
\end{center}
\end{figure}

To determine the red dwarf spectral type, we first subtract our best--fitting white dwarf model from the average of spectra taken between phases 0.8 and 0.2 and fit the red dwarf model spectra to the residuals. The fit is not optimal, and spectral types M3--M5 all fit equally well, with a M4.4 spectrum giving a $\mathrm{\chi_{red}^2=1.85}$ as the best fit, at a distance of $d_{\rm RD, M4}=\mathrm{1303^{+87}_{-75}}$\,pc (see Fig. \ref{fig:Red_spectrafit}). The best--fitting distance for spectral types M3 and M5 are $d_{\rm RD,M3} = \mathrm{1523^{+82}_{-71}}$\,pc and $d_{\rm RD,M5} = \mathrm{783^{+87}_{-75}}$\,pc. Since the fit is not optimal, we adopt an uncertainty of 1 spectral type, M$\mathrm{3--5}$. The distance range associated with this is $\mathrm{1523--783}$\,pc. If we assume that the red dwarf is at the same distance as the white dwarf, the red dwarf has to be over luminous by a factor of 2--7 for spectral types M5--M3. Or, conversely, the surface gravity of the white dwarf has to be 0.18 dex lower if we fix the white dwarf distance to the red dwarf distance (for a fixed white dwarf temperature of 25700\,K).

We also fit the red dwarf models to the interloper spectrum, with a best fit of an M3.1 model at a distance of $\mathrm{1132^{+42}_{-36}}$ pc. This fit is good, $\mathrm{\chi_{red}^2=1.04}$, see Fig. \ref{fig:Red_spectrafit}. A spectral type of M3 is clearly better than either an M4 or an M2 spectrum and the statistical uncertainty is much lower than 0.1 spectral types. Therefore, we conclude that the spectral type of the interloper is M3. 

The spectrum of the interloper is only calibrated photometrically with a standard star observed at the end of the spectroscopic run. The photometric calibration of the spectra of PTF0857 taken just before showed scatter of about 5\%. Using the spectrum, we calculated the magnitude in the $r$ and $i$ bands: $r=19.93\pm0.06$ and $r=18.91\pm0.06$. These measurements are consistent with the values in Table \ref{tab:mags}. If we also take into account the calibration uncertainty, the distance to the interloper is $d_{\rm interloper} = \mathrm{1132^{+96}_{-76}}$\,pc

\subsection{Radial velocities}\label{subsec:rv}
\begin{figure*}
  \begin{center}
    \includegraphics[width=0.49\textwidth]{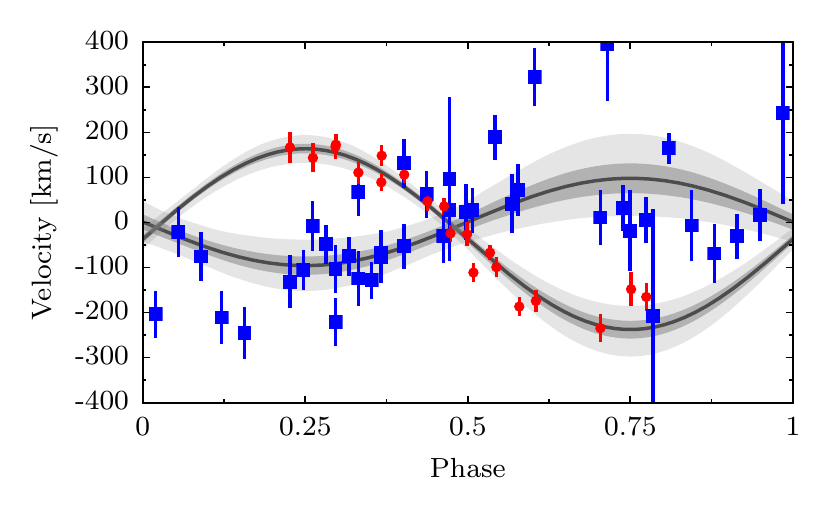}
    \includegraphics[width=0.49\textwidth]{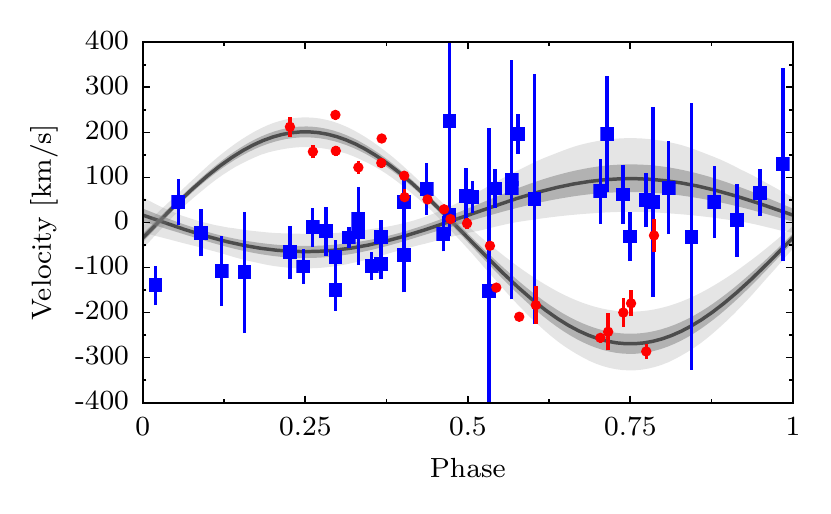}\\
  \caption{Measurements of the radial velocities of the red dwarf as measured from H$\alpha$ (circles) and the white dwarf (squares). The grey bands show the 1 standard deviation and 3 standard deviations of the fits, determined using the bootstrapping procedure. (\emph{Left}) The radial velocities as measured using line fitting. (\emph{Right}) The data and best fit to the radial velocities obtained using cross-correlation.}
  \label{fig:RV_both}
  \end{center}
\end{figure*}

We determined the radial velocity curves from the spectra using the program {\sc molly}. First, we normalized the spectra by fitting a low-order polynomial to the continuum of the spectra. We fit the H$\alpha$ emission line using a single Gaussian with a fixed width but variable height and central wavelength to determine the radial velocity shifts of the red dwarf. By doing this we do ignore the H$\alpha$ absorption line from the white dwarf, but the emission line is much stronger than the absorption line and should not affect the fit to the emission line (see Fig. \ref{fig:averagespectrum}).

The other Balmer lines are dominated by the broad absorption lines originating in the white dwarf's atmosphere, but emission from the secondary is clearly detectable (see Fig. \ref{fig:averagespectrum}). To determine both the white dwarf and red dwarf radial velocity, we fit the H$\beta$, H$\gamma$ and H$\delta$ absorption and emission features simultaneously. The absorption lines were approximated using Gaussian profiles, all with the same radial velocity offset to the laboratory rest frame. For the emission features we also used a Gaussian profile, but allowed for individual offsets to the rest frame. The depth and width of the absorption features and the FWHM of the emission features were kept fixed (determined from the average of spectra between phases 0.4 and 0.6), but we kept the depth and height of the profiles as free parameters. We fit a sinusoid to the phase-folded radial velocity variations, using the period and phase obtained from the photometry and kept the amplitude and zero--point as free parameters. We only used measurements between phases 0.8 and 0.2. A 5$\sigma$ clipping of outliers was applied in two iterations. 
The average red dwarf radial velocity amplitude from the H$\alpha$, H$\beta$, H$\gamma$, and H$\delta$ is $K_\mathrm{RD,obs}=212\pm10\,  \mathrm{km\, s^{-1}}$. This measured radial velocity of the centre of light of the star can still include a $K$--correction, due to the strong irradiation by the white dwarf (see Section \ref{subsec:syspars}). For the white dwarf we measure a radial velocity amplitude of $K_\mathrm{WD}=97\pm22\,  \mathrm{km\, s^{-1}}$. We determined the uncertainties of our fit using a bootstrap method, shown in Fig.~\ref{fig:RV_both}.

To confirm these measurements, we used a cross-correlation of the individual spectra with the average spectrum, using the {\sc rv/fxcor} package in {\sc iraf}. To determine the radial velocity of the white dwarf, the Balmer H$\beta$, H$\gamma$, H$\delta$ and H$\epsilon$ absorption lines were used with the central 12 \AA\ masked. For the red dwarf radial velocity only the H$\alpha$ emission line was used (without any correction of white dwarf absorption). We fitted the resulting radial velocity curve as above, and determined an average red dwarf radial velocity amplitude $K_\mathrm{RD,obs}=235\pm15\,  \mathrm{km\, s^{-1}}$ and $K_\mathrm{WD}=81\pm18\, \mathrm{km\, s^{-1}}$ for the white dwarf, consistent with the first method. The results are shown in Fig.~\ref{fig:RV_both}.

We adopt the values from our first method, $K_\mathrm{RD,obs}=212\pm10\, \mathrm{km\, s^{-1}}$ and $K_\mathrm{WD}=97\pm22\, \mathrm{km\, s^{-1}}$. The formal uncertainty of this method is larger for the white dwarf radial velocity amplitude, but this method is the least likely to be affected by blending of the features, because both absorption and emission profiles are fitted at the same time. The uncertainty on the radial velocity of the white dwarf primary, $\mathrm{\Delta} K_{\rm WD}/K_{\rm WD}$, is 0.22. This uncertainty is reflected in the mass-ratio [$q=M_{\rm RD}/M_{\rm WD}=K_{\rm WD}/K_{\rm RD}$], see Section \ref{subsec:LC}).

\subsection{Light curves}\label{subsec:LC}
\begin{figure*}
\begin{center}
  \includegraphics[width=\textwidth]{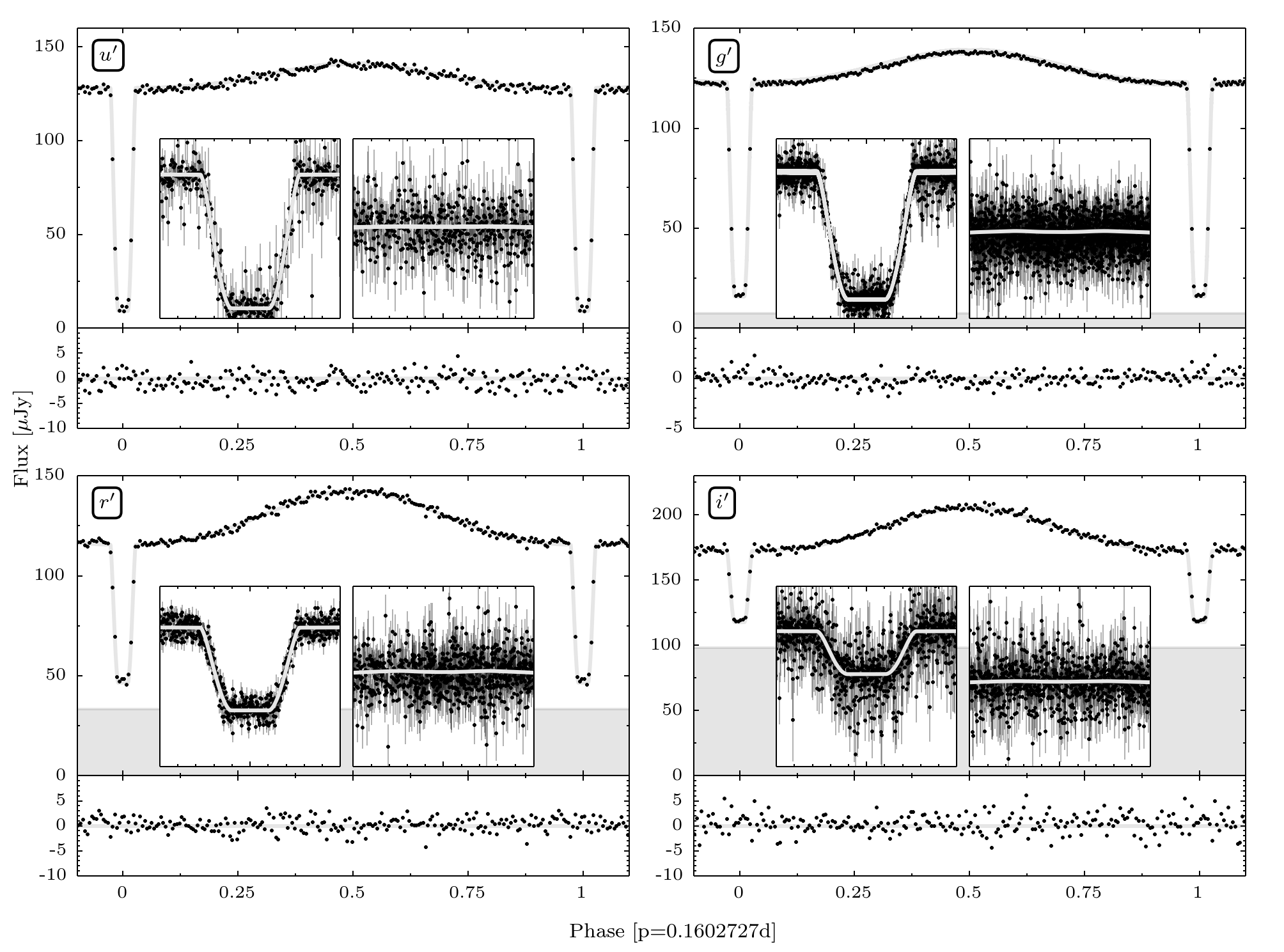}
  \caption{The folded and binned ULTRACAM light curves and best fitting models in $u^\prime$, $g^\prime$, $r^\prime$ and $i^\prime$. The main panels show the binned data and the model. The greyed out area indicates the contribution of the interloper star. The bottom panels shows the residuals of the fit. The insets show the unbinned data and the models near the primary and secondary eclipses.}
 \label{fig:LC_models}
\end{center}
\end{figure*}

By modelling the light curves we can put strong constraints on the system parameters. To construct a model light curve given a set of binary star parameters, we use {\sc lcurve} (written by T.R. Marsh et al., see \citealt{2010MNRAS.402.1824C}). The {\sc lcurve} code uses grids of points to model the two stars. The shape of the stars in the binary is set by a Roche potential for a certain mass-ratio. {\sc lcurve} assumes co-rotation and zero orbital eccentricity.

The amount of light each point on the surface grid emits depends on a number of parameters, the main one being the star's effective temperature. We approximate the spectral energy distribution of both stars with a blackbody and calculate the flux at the effective wavelength of the bandpass. {\sc lcurve} takes limb darkening into account. For the red dwarf we use a quadratic limb darkening law with parameters for a $3000\,$K red dwarf \citep{Claret:2011}. For the white dwarf we use a Claret limb darkening prescription with the parameters for a 25000K, $\log g$=7.5 white dwarf taken from \citet{2013ApJ...766....3G}. In addition, we also take into account gravity darkening for the red dwarf. We use the relation between the intensity and surface gravity: $I\propto g^y$, with $y$ a free parameter between 0.1 and 1.2 (see \citealt{2013ApJ...766....3G}).

Since the white dwarf is very hot and close to the secondary, irradiation of the secondary by the white dwarf cannot be neglected. A back-of-the-envelope calculation shows that the irradiated energy flux can be a factor 200 higher than the energy flux from the red dwarf. This `reflection' effect is a combination of heating and reflection, both contributing to the flux emitted by the secondary. In our model, we approximate this effect using an albedo factor, which determines the fraction of the light that is absorbed. This will locally increase the temperature and therefore increase the flux emitted by that area on the surface. The fraction of light that is not absorbed is ignored in the model. In theory the albedo should be between 0 and 1, but this method is a very crude approximation for the reflection effect, and thus we do not constrain the range of the albedo parameter.

We combine our {\sc lcurve} model with a Monte Carlo Markov Chain method, as implemented in the package {\sc emcee} \citep{2013PASP..125..306F} to explore the solution space. We calculate the solution for the light curves per passband, using as priors $T_{\mathrm{WD}}=25700\pm400$\,K and the contribution from the interloper as measured from the ULTRACAM images (Table \ref{tab:mags}). When inspecting the residuals we find a trend over time, likely due to a difference in airmass. To remove the trends in the data, we add a low second order polynomial function to our model, and use this to correct the data. Besides the long scale trend, we also noticed some minor flaring behaviour, simultaneously in the $g^{\prime}$ band and $r^{\prime}$ band data, possibly from chromospheric activity of the red dwarf. We removed the affected data points from our light curve as studying the flare is not within the scope of this paper.

We first fitted the data without any constraint on the mass ratio. The best fitting models to the individual light curves are shown in Fig. \ref{fig:LC_models}. No systematic deviations can be seen in the residuals. The primary eclipse is fitted well in all bands. The best fitting models do not show any indication of a secondary eclipse in any of the bands. 

After a more detailed inspection of the parameter uncertainties, we deduced that the mass ratio is only very marginally constrained by the light curve. The effect of the mass-ratio on the light curve is in the amount of ellipsoidal variation, which is a very small effect compared to the reflection effect. There is a small preference for models with a mass ratio of $q\approx 0.1-0.2$, but the $\Delta\chi^2$ improvement compared to $q=0.5$ models is only 35 out of 15337. Statistically, this means that a mass ratio of $q=0.5$ is ruled out, but this assumes that our model and data are both perfect, especially in how we treat limb darkening. This not the case, and we therefore decide to treat the mass-ratio as a systematic uncertainty and calculate a solution for a range of mass-ratios (0.1, 0.2, 0.25, 0.3, 0.35, 0.4, 0.5).

\subsection{System parameters}\label{subsec:syspars}
We determine the mass of both components using the mass function, 
\begin{equation}
  M_{\rm A} = \dfrac{(K_{\rm WD}+K_{\rm RD})^2 K_{\rm B} P}{2 \pi G \sin(i)^3}
\end{equation}
with $A$ and $B$ being either the white dwarf (WD) or the red dwarf (RD), $M$ the mass, $K$ the radial velocity amplitude, $P$ the orbital period and $i$ the inclination. Once the masses are known, we can calculate the orbital separation ($a$) using Kepler's third law. 

The light curve model gives us very good constraints on the relative sizes of the components and the inclination ($i$), but we need spectroscopy to determine the absolute masses of both components. For many binaries it is straightforward to obtain phased resolved spectroscopy and use periodic velocity shifts in spectral features to obtain the radial velocities.
For PTF0857 there are two complications. One is the strong reflection effect, which shifts the centre--of--light of the red dwarf away from its centre-of-mass. This means that the measured radial velocity of the red dwarf is lower than the centre--of--mass radial velocity. We describe this effect using the following equation (first used by \citealt{Wade:1988}):
\begin{equation}\label{eq:RV_cor}
  K_{\mathrm{RD}} = K_{\mathrm{RD},\mathrm{obs}}/({1-f\ r_{2,\mathrm{L1}}(1+q))},
\end{equation}
with $K_{\mathrm{RD},\mathrm{obs}}$ the observed radial velocity amplitude of the irradiated component (the red dwarf in the case of PTF0857), $f$ a value between 0 and 1 that indicates the offset of the centre--of--light to the centre-of-mass of the secondary, and $r_{\rm RD,\mathrm{L1}}$ the radius of the red dwarf in the direction of the inner Lagrangian point, divided by the orbital semi-major axis, i.e. $r_{\rm RD, L1} = R_{\rm RD,L1}/a$. This adds an additional parameter to our model, $f$, which lies, by definition, between 0 and 1, but has a fixed dependence on the geometry.  We use the Roche geometry (set by the mass ratio $q$) and the relative size of the red dwarf from the light--curve models to calculate $f$. This shows that the value of $f$ is 0.69 for a mass-ratio of 0.3, and only changes by 0.03 for mass ratios of $\Delta q=0.1$. Therefore, we decide to use a range of $f=0.66--0.72$ for all models. See Appendix \ref{appendix} for the calculation of, and a further discussion on, the expected value for $f$.

A second complication is that the spectral features of the white dwarf and red dwarf are blended, as outlined in Section \ref{subsec:rv}, and therefore the uncertainty on the radial velocity measurement of the white dwarf is high. In combination with the high signal-to-noise light curve, the uncertainty in $K_\mathrm{WD}$ (and therefore the mass-ratio $q$) will dominate the error budget. To better understand the source of uncertainties in the different parameters, we calculate both a statistical uncertainty on the model parameters, as well as how the uncertainty in $K_\mathrm{WD}$ propagates through. We choose the models with mass ratios that are most compatible with the observed value of $K_{\mathrm{WD}}=97\,\mathrm{km\,s^{-1}}$, and derive how each parameter changes if $K_{\mathrm{WD}}$ is one standard deviation lower or higher, $K_{\mathrm{WD}}=75\, \mathrm{km\,s^{-1}}$ and $K_{\mathrm{WD}}=119\ \mathrm{km\,s^{-1}}$. We use linear interpolation between models with different mass ratios to obtain these values.

The result of the combined fitting for the system parameters is given in Table \ref{tab:LC_results}. The light curve parameters are fit independently per band, except for the mass-ratio ($q$), which we determined using $K_\mathrm{WD}$, $K_\mathrm{RD}$, $f$ and $r_\mathrm{RD}$.
The light curves in the four different bands essentially give us four independent measurements of the light curve parameters (see Table \ref{tab:LC_results}). Most of the parameters are consistent with each other, but there are a few inconsistencies. First, the temperature of the secondary is significantly higher as measured in the $u^\prime$ band than in the other bands. This is possibly a result of using a blackbody approximation to calculate the flux \citep[this is also seen in][]{Parsons:2009}, but could also indicate excess flux at shorter wavelengths due to the irradiation \citep{2004ApJ...614..338B}. The red dwarf temperatures ($T_\mathrm{RD}$) are also not entirely consistent for the $g^\prime$, $r^\prime$, and $i^\prime$ bands. This could be due to systematic uncertainties in determination of the contribution by the interloper, which was already noted in Section \ref{subsec:mags}, or the fact that also the spectrum of a low-mass star is not a blackbody, certainly in the redder parts of the spectrum where strong TiO absorption comes in. Related to the inconsistently high red dwarf temperature is the unphysically high albedo (>1) in the $u^\prime$ band results.

A second inconsistency are the values for $r_\mathrm{WD}$, $r_\mathrm{RD}$, and $i$ for the $i^\prime$ band models compared to the results from the other bands. These three parameters are correlated and set by the duration of the ingress and egress, and the eclipse. The difference is $\unsim2-4$ standard deviations between the $i^\prime$ and $g^\prime$ solutions. This could be caused by the fact that in the $i^\prime$ band the interloper outshines the eclipsing binary PTF0857, which can cause systematic uncertainties. Because the solution is consistent for the $u^\prime$, $g^\prime$, and $r^\prime$ bands, we choose to accept those solutions, and we will ignore the solution in the $i^\prime$ band in further discussions.

\begin{table*}
  \centering
 \caption{Results of the fits to the ULTRACAM light curves in the four different bands. We use linear interpolation to calculate the mass ratio for which the white dwarf radial velocity matches the observed $97\,\mathrm{km\ s^{-1}}$. We calculate the value and uncertainty of each parameter for that mass ratio. For each parameter two uncertainties are given (both indicate the 16\%--84\% interval, equivalent to 1 standard deviation); the first is the uncertainty as a result of the uncertainty on the mass ratio $q$, the second is the statistical uncertainty. The zero--phase time is 55957.121914 BMJD (TDB), the best--fitting value for the $g^\prime$ band, and the table gives $\Delta t_0$, the offset from this value. A $\textsuperscript{p}$ indicates that we used a prior of some sort for that parameter, see text for details.}
 \def\arraystretch{1.2}
 \begin{tabular}{ll|llll} \label{tab:LC_results}
  & & $u^\prime$ & $g^\prime$ & $r^\prime$ & $i^\prime$  \\
  \hline
  \hline
\multicolumn{6}{l}{\bf{Light curve parameters}}   \\
Mass ratio & $q$ & $0.31_{-0.06}^{0.06}$ & $0.31_{-0.06}^{0.06}$ & $0.32_{-0.06}^{0.05}$ & $0.34_{-0.06}^{0.06}$ \\
Phase 0 offset & $\Delta t_0\ (10^{-6}\, \mathrm{d})$ & $-11^{+0\ +9}_{-1\ -9}$ & $0^{+0\ +3}_{+0\ -3}$ & $2^{-0\ +11}_{-1\ -11}$ & $8^{-4\ +23}_{-1\ -25}$ \\
WD effective temperature\textsuperscript{p} & $T_1\ \mathrm{(K)}$ & $25700^{-0\ +400}_{+0\ -400}$ & $25700^{-0\ +400}_{-0\ -400}$ & $25700^{-0\ +400}_{-0\ -400}$ & $25700^{-0\ +400}_{+0\ -400}$ \\
RD effective temperature & $T_2\ \mathrm{(K)}$ & $4200^{-70\ +250}_{+60\ -220}$ & $3280^{-50\ +60}_{+90\ -60}$ & $3110^{-90\ +80}_{+120\ -90}$ & $3290^{-150\ +210}_{+230\ -190}$ \\
WD radius/$a$ & $r_\mathrm{WD} $ & $0.0199^{-0.0011\ +0.0010}_{+0.0014\ -0.0007}$ & $0.0212^{-0.0012\ +0.0003}_{+0.0017\ -0.0002}$ & $0.0205^{-0.0012\ +0.0009}_{+0.0016\ -0.0008}$ & $0.0275^{-0.0020\ +0.0025}_{+0.0029\ -0.0022}$ \\
RD radius/$a$ & $r_\mathrm{RD,L_1}$ & $0.352^{+0.017\ +0.022}_{-0.019\ -0.038}$ & $0.345^{+0.016\ +0.010}_{-0.023\ -0.010}$ & $0.339^{+0.020\ +0.014}_{-0.024\ -0.015}$ & $0.285^{+0.022\ +0.020}_{-0.025\ -0.020}$ \\
RD radius/$a$ & $r_\mathrm{RD}$ & $0.281^{+0.012\ +0.003}_{-0.016\ -0.012}$ & $0.279^{+0.013\ +0.002}_{-0.017\ -0.003}$ & $0.277^{+0.014\ +0.003}_{-0.017\ -0.004}$ & $0.256^{+0.017\ +0.012}_{-0.020\ -0.012}$ \\
Binary inclination & $i\ \mathrm{(^\circ)}$ & $76.5^{-0.8\ +0.6}_{+1.0\ -0.1}$ & $76.5^{-0.8\ +0.1}_{+1.1\ -0.1}$ & $76.6^{-0.9\ +0.3}_{+1.1\ -0.2}$ & $77.7^{-1.1\ +0.8}_{+1.3\ -0.6}$ \\
albedo &  & $1.91^{-0.04\ +0.23}_{+0.09\ -0.21}$ & $0.96^{-0.03\ +0.05}_{+0.04\ -0.05}$ & $0.89^{-0.04\ +0.05}_{+0.07\ -0.04}$ & $0.62^{-0.07\ +0.06}_{+0.12\ -0.05}$ \\
Interloper contribution$^P$ & $L_\mathrm{3}\ \mathrm{(\mu Jy)}$ & $3.8^{-0.3\ +2.1}_{+0.6\ -2.6}$ & $11.2^{-0.3\ +0.5}_{+0.1\ -0.6}$ & $36.9^{-0.1\ +1.2}_{+0.1\ -1.3}$ & $97.8^{-0.1\ +2.9}_{+0.1\ -2.9}$ \\
RD gravity darkening & $\gamma_\mathrm{2}$ & $0.7^{+0.0\ +0.4}_{-0.0\ -0.4}$ & $1.1^{+0.0\ +0.1}_{-0.0\ -0.2}$ & $0.8^{+0.0\ +0.3}_{-0.1\ -0.3}$ & $0.9^{-0.0\ +0.2}_{-0.0\ -0.4}$ \\
  \hline
\multicolumn{6}{l}{\bf{System parameters}}   \\
Semi--major axis & $a\ \mathrm{(R_{\odot})}$ & $0.88^{+0.08\ +0.05}_{-0.08\ -0.06}$ & $0.87^{+0.08\ +0.04}_{-0.08\ -0.04}$ & $0.87^{+0.08\ +0.05}_{-0.08\ -0.04}$ & $0.83^{+0.08\ +0.05}_{-0.08\ -0.05}$ \\
WD mass & $M_\mathrm{WD}\ \mathrm{(M_{\odot})}$ & $0.61^{+0.15\ +0.12}_{-0.13\ -0.11}$ & $0.61^{+0.15\ +0.10}_{-0.14\ -0.09}$ & $0.60^{+0.16\ +0.10}_{-0.13\ -0.09}$ & $0.50^{+0.13\ +0.09}_{-0.11\ -0.08}$ \\
RD mass & $M_\mathrm{RD}\ \mathrm{(M_{\odot})}$ & $0.19^{+0.09\ +0.04}_{-0.07\ -0.04}$ & $0.19^{+0.09\ +0.03}_{-0.07\ -0.03}$ & $0.19^{+0.09\ +0.03}_{-0.07\ -0.03}$ & $0.17^{+0.08\ +0.03}_{-0.06\ -0.03}$ \\
WD radius & $R_\mathrm{WD}\ \mathrm{(R_{\odot})}$ & $0.0175^{+0.0006\ +0.0010}_{-0.0004\ -0.0010}$ & $0.0185^{+0.0006\ +0.0009}_{-0.0004\ -0.0009}$ & $0.0178^{+0.0005\ +0.0011}_{-0.0004\ -0.0011}$ & $0.0227^{+0.0004\ +0.0021}_{-0.0000\ -0.0019}$ \\
RD radius & $R_\mathrm{RD}\ \mathrm{(R_{\odot})}$ & $0.24^{+0.03\ +0.02}_{-0.04\ -0.02}$ & $0.24^{+0.03\ +0.01}_{-0.04\ -0.01}$ & $0.24^{+0.04\ +0.01}_{-0.04\ -0.01}$ & $0.21^{+0.04\ +0.02}_{-0.04\ -0.02}$ \\
WD surface gravity & $\log{g}$ & $7.74^{+0.07\ +0.04}_{-0.08\ -0.06}$ & $7.69^{+0.07\ +0.03}_{-0.09\ -0.03}$ & $7.71^{+0.08\ +0.04}_{-0.09\ -0.04}$ & $7.43^{+0.09\ +0.08}_{-0.11\ -0.09}$ \\
WD radial velocity\textsuperscript{p} & $K_\mathrm{WD}\ \mathrm{[km\ s^{-1}]}$ & $97^{+22\ +6}_{-22\ -6}$ & $97^{+22\ +5}_{-22\ -5}$ & $97^{+22\ +5}_{-22\ -5}$ & $97^{+22\ +5}_{-22\ -5}$ \\
RD radial velocity\textsuperscript{p} & $K_\mathrm{RD}\ \mathrm{[km\ s^{-1}]}$ & $309^{+14\ +19}_{-13\ -19}$ & $309^{+13\ +16}_{-15\ -16}$ & $306^{+15\ +16}_{-15\ -16}$ & $288^{+13\ +16}_{-13\ -16}$ \\
Roche lobe fill factor &   &  $0.987^{+0.001\ +0.009}_{-0.001\ -0.041}$ & $0.982^{+0.002\ +0.007}_{-0.006\ -0.010}$ & $0.974^{+0.007\ +0.012}_{-0.008\ -0.016}$ & $0.883^{+0.019\ +0.042}_{-0.022\ -0.041}$ \\
   \end{tabular}
\end{table*}

\section{Discussion}\label{sec:discussion}

\begin{figure*}
  \begin{center}
    \includegraphics[width=0.99\textwidth]{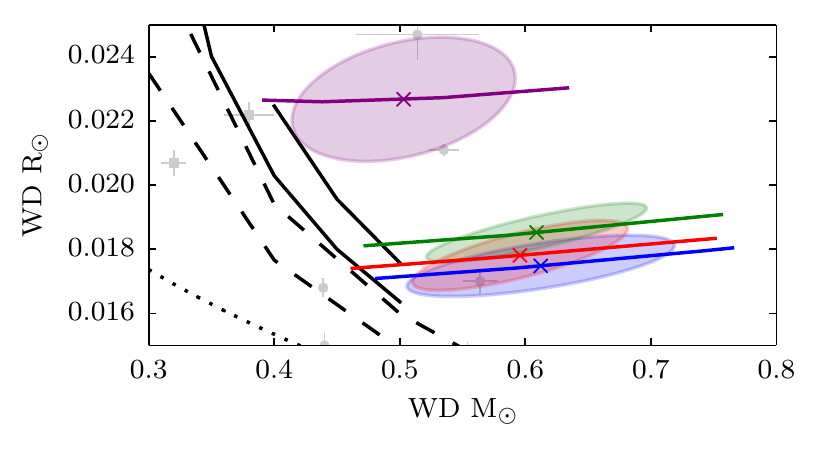}
    \includegraphics[width=0.99\textwidth]{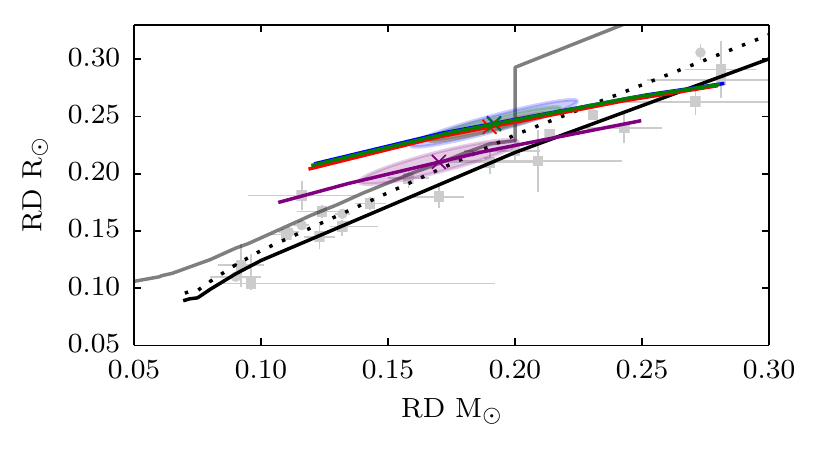}\\
  \caption{The mass versus radius of the white dwarf (top) and red dwarf (bottom). The 'x' markers show the most probable solutions for the masses and radii for the fit to the $u^\prime$,$g^\prime$,$r^\prime$, and $i^\prime$ light curves (blue, green, red and purple). The ellipses indicate the statistical uncertainty on the solution, and the solid coloured lines show the range of solutions due to the uncertainty in the mass ratio. (\emph{Top}) The dotted line indicates the zero-temperature mass-radius relation by \citet{1988ApJ...332..193V}, the dashed lines show the models for a 25000K CO white dwarf with a thin and thick hydrogen layer ($q_H=10^{-10}$, $q_H=10^{-4}$), taken from \citet{2001PASP..113..409F}, and the black line shows a model of a 25000K He white dwarf from \citet{1999MNRAS.303...30B}. The grey circles indicate the mass and radius of white dwarfs in other PCEB binaries \citep{2010MNRAS.407.2362P,2012MNRAS.420.3281P,2012MNRAS.426.1950P,2016MNRAS.458.2793P}. Grey squares indicate the mass and radius from the two white dwarfs in the eclipsing binary CSS 41177 \citep{2014MNRAS.438.3399B}.
  (\emph{Bottom}) The solid black line represents a 5Gyr isochrone by \citet{2015A&A...577A..42B}, the dashed line shows the same isochrone, but with the radius increased by 7\% as predicted for irradiated red dwarfs, and the grey line indicates the mass-radius relation for red dwarfs in cataclysmic variables \citep{Knigge:2011}. Grey circles indicate the mass and radius of red dwarfs in other PCEB binaries \citep{2010MNRAS.407.2362P,2012MNRAS.420.3281P,2012MNRAS.426.1950P,2016MNRAS.458.2793P}. Grey squares indicate the mass and radius of single red dwarfs or red dwarfs with a main-sequence companion from \citealp{2007ApJ...663..573B,2007ApJ...660..732L,2013MNRAS.431.3240N,2013A&A...553A..30T,2014MNRAS.437.2831Z}.}
  \label{fig:MR_both}
  \end{center}
\end{figure*}

\subsection{Mass, Radius and Temperature}
The masses and radii of the two components are plotted in Fig. \ref{fig:MR_both}. As can be seen in this figure and Table \ref{tab:LC_results}, the uncertainty on the mass of the white dwarf is dominated by the uncertainty on the mass ratio. For a mass ratio in the range $q=0.25--0.38$, the white dwarf in PTF0857 has a mass between $M_\mathrm{{WD}}=0.47--0.71\,\mathrm{M_\odot}$, with a statistical uncertainty of $\unsim0.10\, \mathrm{M_\odot}$.

The mass and radius of the white dwarf are both consistent with He-core white dwarf models and CO-core white dwarf models, as can be seen in Fig. \ref{fig:MR_both}. The models of $25000$\,K He white dwarfs \citep{1999MNRAS.303...30B} match the solution in the mass range of $M_\mathrm{{WD}}=0.45--0.50\,\mathrm{M_\odot}$.  Models for $25000$\,K CO-core white dwarfs \citep{2001PASP..113..409F} have a slightly smaller radius for a given mass compared to the He white dwarf models. Solutions with a mass in the range of $M_\mathrm{{WD}}=0.42--0.45\,\mathrm{M_\odot}$ match these models. Current uncertainties on the white dwarf mass and radius exclude a distinction on the white dwarf core composition, but both solutions are at the lower end of the mass range, which corresponds with low mass ratios in the binary system.

The red dwarf in PTF0857 has a mass between $M_\mathrm{{RD}}=0.12--0.28\,\mathrm{M_\odot}$ for $q=0.25--0.38$, with a statistical uncertainty of $\unsim0.03\,\mathrm{M_\odot}$. The total uncertainty on the mass and radius of the red dwarf is even more dominated by the uncertainty in the mass ratio. If we compare our solution space to mass--radius models for single red dwarfs \citep[solid black line in Fig.\ \ref{fig:MR_both}]{2015A&A...577A..42B}, the solution with a mass of $0.27\,\mathrm{M_\odot}$ and at a high value of $q$ fits best. The model temperature for stars of this mass is $\approx$3340\,K, consistent with results from our light curve modelling. Red dwarfs of this mass typically have spectral type of $\approx$M3.5 \citep{2013A&A...556A..15R}, which agrees with our measurement in Section \ref{subsec:spectemp}, M3-M5. 

There are some caveats in using models for single M-dwarfs. \citet{2000A&A...360..969R} showed that for M dwarfs that are strongly irradiated, such as PTF0857, the radius of the red dwarf can increase by about 7 percent (indicated by the dashed line in Fig. \ref{fig:MR_both}). A second effect of being part of a binary is that the red dwarf has a short rotation period making it mostly likely tidally synchronised to the orbital period. This short rotational period can make the M dwarf very magnetically active. High stellar activity inhibits convection, and for the star to remain in hydrostatic equilibrium, it has to increase its radius \citep{2006Ap&SS.304...89R,2007A&A...472L..17C}. If either effect has increased the equilibrium radius of the red dwarf in PTF0857 a lower mass is required for it to fit the current solutions. For slightly oversized (`bloated') red dwarfs, our solutions intersect at $0.22\,\mathrm{M_\odot}$ (black dashed line in Fig.\ \ref{fig:MR_both}). Fig.\ \ref{fig:MR_both} shows that, given the current uncertainty on the mass ratio, we cannot make a distinction between a bloated and a normal red dwarf radius. 

The derived ranges on the masses, radii and temperatures of both the white dwarf and the red dwarf are consistent with stellar structure models. Given the derived radii, the stellar structure models would predict lower than measured masses. In comparing our solutions with varying mass ratios this points towards mass-ratios at the middle to higher end of the allowed range. 

\subsection{Surface gravity, spectral type, and distance}
We calculated the distance to the system when fitting the model spectra to the data (Section \ref{subsec:spectemp}). There is a slight discrepancy between the surface gravity of the white dwarf as measured by fitting the spectra, $\log{g}=7.86\pm0.07$, and as derived from the white dwarf mass and radius in the light curve modelling $\log{g}=7.69^{+0.07\ +0.03}_{-0.09\ -0.03}$ (for the $g^\prime$ band). If the actual mass ratio is on the lower end of the allowed range, as discussed in the previous paragraph, the discrepancy between the spectra and the light curve modelling would grow.  

The discrepancy in the surface gravity is likely due to systematic errors in fitting the spectra. The surface gravity from the spectra is mainly determined by the shape of the Balmer absorption lines of the white dwarf. These are contaminated by the red dwarf line emission. Considering this, we adopt the surface gravity from the light curve fit as the `true' surface gravity of the white dwarf in PTF0857.

Because the surface gravity and distance are strongly correlated, we need to correct the distance determination to PTF0857. If we assume the surface gravity of the white dwarf given by the light curve results, the earlier distance estimate to the white dwarf based on the spectra, needs to be increased by a factor of $\mathrm{1.21}$ to $d=\mathrm{686^{+128\ +23}_{-50\ -21}}$\,pc (the first uncertainties are the systematic uncertainties due to mass-ratio and second the statistical uncertainties), which we adopt as the distance to PTF0857. 

The distance range associated with a red dwarf of spectral type M3-M5 is $\mathrm{1523-783}\,$pc. The lower end of this distance estimate (for an M5 red dwarf) is barely consistent with the upper end of the white dwarf distance range. This is not uncommon; \citet{2007MNRAS.382.1377R} compared distance estimates from white dwarfs and red dwarfs for white dwarf--red dwarf binaries and found that red dwarf distance estimates are often higher than the white dwarf distances. \citet{2007MNRAS.382.1377R} discuss a number of causes, including systematic problems in fitting the white dwarf or systematic problems in determining the red dwarf spectral type, problems with the spectral type - radius relation, effects due to close binarity, and effects due to age. If the red dwarf is oversized for its mass this would help to alleviate the distance discrepancy.

From the single spectrum of the interloper, we determined the spectral type M3 at a distance of $\mathrm{1132^{+96}_{-76}}\,$pc. The spectrum and SDSS-colour match the spectral type, so this measurement is robust. This distance indicates that the interloper is likely a background object if we assume the white dwarf distance is the distance to the binary system. We compared images from the DSS (Digitized Sky Survey) obtained in 1955 with recent SDSS images. The PSF of the DSS image is not good enough to distinguish PTF0857 from the interloper, but the angular distance between PTF0857 and the interloper does not seem to have changed. There is also no significant difference in the relative position of the blended source compared to field stars in the SDSS images, so neither of the objects has a large (relative) proper motion.

\section{System evolution}\label{sec:evolution}
There are two possible scenarios for the formation history of PTF0857; either it emerged from the common envelope as we see it now, or it is a CV that is currently detached (dCV) and is crossing the period gap. It is difficult to distinguish between these two scenarios as the system parameters of dCVs are the same as a subset of the PCEBs.

If PTF0857 is a PCEB that has emerged from the common-envelope phase, the system would not have changed much since emerging. Given its current temperature, we estimate white dwarf cooling age $\tau_{\rm WD}\unsim25\,$Myr, if it is a CO white dwarf \citep{Wood:1995}, or $\tau_{\rm WD}\unsim50\,$Myr for a helium white dwarf \citep{2007MNRAS.382..779P}. This would also be the time since it emerged from the common-envelope. If the system is losing angular momentum due to gravitational waves only, the orbital period decreased just $\unsim$1-2 minutes since then, which means it emerged from the common-envelope right in the orbital period gap of cataclysmic variables. 
Alternatively the system could have gone through an sdB phase (lasting about $\unsim150\,$Myr) after emerging from the common envelope. This scenario requires that the white dwarf mass is around $0.48\,\mathrm{M_\odot}$, which is at the lower mass bound of the solution space. It also means that gravitational wave radiation has had a longer time to shrink the system, which implies a common-envelope-exit at a period that is slightly longer, but no more than $\approx$8\,min, depending on the lifetime of the shell burning in the sdB star.

If PTF0857 is a dCV, the system parameters should be similar to CVs just above and below the period gap and consistent with CV evolution models. Measured white dwarf temperatures for CVs above the gap range from 15\,000K to 50\,000K \citep{2009ApJ...693.1007T,2017MNRAS.466.2855P}, while the CV evolution models predict temperatures in the range of 23\,000-30\,000K at the upper end of the period gap. The white dwarf in PTF0857 (25\,700K) fits within this range, but it should have cooled down after entering the period gap. Under the emission of gravitational wave radiation only it will have taken $\approx$0.8 Gyr to decrease the orbital period to its current value. The cooling age of the white dwarf is much shorter than this and therefore the system is unlikely to have entered the period gap at $\approx$3.18 hours. The actual temperature of the white dwarf at the entry into the orbital period gap is not relevant for this discrepancy between the cooling time and the gravitational in-spiral time. Fig.~5 in \citet{2009ApJ...693.1007T} indicates that very hot WDs in CVs at the upper end of the period gap do exist, but after close to a billion years they should all have cooled down to values far lower than 25\,700\,K. In a recent paper, \cite{2016arXiv160107785Z} used binary population models and numerical simulations to predict the system parameters of the observed population of PCEBs and dCVs. They show that white dwarfs with a temperature of 25\,700K occur in both PCEB and dCV systems with an orbital period of 2.5\,h, and conclude that the white dwarf temperature cannot be used to distinguish if a system is a dCV or PCEB. It is however not clear how these systems would `escape' the cooling age argument given above. The same study does show that the distribution of the white dwarf masses is different for the two populations; massive white dwarfs ($M>0.8\, \mathrm{M_{\odot}}$) only occur in dCV systems. The mass of the white dwarf in PTF0857 is $M_\mathrm{WD}=0.61^{+0.15}_{-0.14}\, \mathrm{M_{\odot}}$, which occurs in both scenarios and is consistent with measured white dwarfs in CVs both above and below the period gap. In conclusion, with the uncertainties in the white dwarf mass and in the evolutionary models we cannot exclude either scenario based on the white dwarf properties, although a detached CV scenario is unlikely given the high temperature of the white dwarf.

The Red dwarf spectral type is consistent with measured spectral types of CVs, as well as the model value of M4.0 \citep{Knigge:2011}. If we compare the mass and radius of the red dwarf to measurements of red dwarfs in CVs above and below the period gap \citep[Table 1 in ][]{2006MNRAS.373..484K}, we note that they are consistent both in mass and radius with CVs below the gap, but fall below the range of masses and radii of CV-red dwarfs above the gap, see also the grey line in Fig.\ \ref{fig:MR_both}. The CV-model by \citet{Knigge:2011}, which is based on measurements, predicts that the mass of a red dwarf that has just stopped mass transfer (a dCV) is $M_\mathrm{RD} = 0.20 \pm 0.02\,\mathrm{M_\odot}$. While the red dwarf radius is still inflated just after mass-transfer stops, it quickly shrinks down to a radius of $R_\mathrm{RD} = 0.23\, \mathrm{R_\odot}$ \citep{1996MNRAS.279..581S}. This combination of mass and radius is within the solution space for PTF0857 (see Fig. \ref{fig:MR_both}). The red dwarf is filling $\unsim$98 percent of the Roche lobe in radius ($\unsim$94\% of the volume), well within the predicted range for dCVs ($>$76\%) for the orbital period of PTF0857. In conclusion: the red dwarf mass and radius are consistent with measurements of CV-red dwarfs at the upper end of the period gap, and also consistent with model values of the red dwarf radius in the dCV scenario.

Regardless of which of the two scenarios is correct, the future of the binary is the same: the binary separation will shrink due to gravitational wave emission and the system will come into contact in $\approx$70\,Myr. If the radius of the secondary does not change, stable mass transfer will start at an orbital period of about $2.47$\,h, very close to its current value, and the system will continue its evolution as a cataclysmic variable, very similar to the known systems SDSS\,J1627+1204 and SDSS\,J0659+2525 (\citealt{2003A&A...404..301R, 2009JBAA..119..144S}). If it is currently a PCEB that emerged from the common envelope at a period close to the current period, it implies that a direct injection of systems to the cataclysmic variable below the period gap is possible and that therefore the space density of systems above the gap should be lower than that of systems below the gap.

\section{Summary}\label{sec:summary}
PTF1 J085713+331843 is an eclipsing binary with an orbital period of 0.1060272(4)\,d, consisting of a 25\,700\,K DA white dwarf and a M3--5 red dwarf. The light curve shows a total primary eclipse of the white dwarf and a strong reflection effect but no secondary eclipse. The system has a nearby, $\unsim1.5\,$arcsec, neighbour with a spectral type M3, most likely a background object.

We analysed high cadence ULTRACAM light curves in the $u^\prime$, $g^\prime$, $r^\prime$ and $i^\prime$ bands and phase-resolved spectroscopy to determine the system parameters. The white dwarf's radial velocity accuracy is the main source of uncertainty on the system parameters. The white dwarf has a derived mass of $M_\mathrm{WD}=0.61^{+0.15}_{-0.14}\, \mathrm{M_{\odot}}$. The white dwarf mass--radius solution is compatible with models of both He and CO white dwarfs. The red dwarf mass is $M_\mathrm{RD}=0.19^{+0.09}_{-0.07}\, \mathrm{M_{\odot}}$, and matches red dwarf mass-radius models. The best solutions to mass--radius models for the white dwarf and red dwarf are consistent which each other within the observational uncertainties. 

To improve our measurements, we require higher signal-to-noise phase-resolved spectroscopy over at least one orbit to measure the white dwarf radial velocity amplitude with higher precision. In addition, the NaI absorption doublet near 8200\,\AA\ can be used to measure the radial velocity of the centre of mass of the red dwarf, although the radial velocity measurements need to be corrected for the irradiation effect, as was done in \citet{Parsons:2009}. 

The semi--major axis of the system is smaller than a solar radius, and therefore the system must have experienced a common-envelope phase in its evolution. Within the current uncertainties we cannot clearly distinguish between a detached CV or a PCEB. In the former case, the system has already been a cataclysmic variable and is currently in hibernation. In the latter case, the system emerged from the common-envelope at an orbital period close to its current period, which then happened about $25--50\,$Myr ago. The system will keep losing angular momentum due to gravitational wave emission and start stable mass transfer in only $\approx$70 Myr, at an orbital period close to its current value. It will become one of the few known cataclysmic variables in the period gap.

\section{Acknowledgements}
We thank the anonymous referee for his/her helpful comments improving this manuscript, and thank T. Kupfer and C. Knigge for many useful discussions.

The results presented in this paper are based on observations collected with ULTRACAM, supported by STFC grants, at the William Herschel Telescope operated on the island of La Palma by the Isaac Newton Group in the Spanish Observatorio del Roque de los Muchachos of the Institutions de Astrofisica de Canarias.

Based on observations of the Palomar Transient Factory. The Palomar Transient Factory is a scientific collaboration between the California Institute of Technology, Columbia University, Las Cumbres Observatory, the Lawrence Berkely National Laboratory, the National Energy Research Scientific Computing Center, the University of Oxford, and the Weizmann Institute of Science.

Based on observations collected at the Palomar Observatory with the $200^{\prime\prime}$ Hale Telescope, operated by the California institute of Technology, its divisions Caltech Optical Observations, the Jet Propulsion Laboratory (operated for NASA) and Cornell University. 

Some of the data presented here were obtained by the Catalina Real-time Transient Survey (CRTS). CRTS are supported by the U.S. National Science Foundation  under  grants  AST-0909182  and  CNS-0540369. The CSS survey is funded by the National Aeronautics and Space Administration under Grant No. NNG05GF22G  issued through the Science Mission Directorate Near-Earth Objects Observations Program.

Funding for the Sloan Digital Sky Survey IV has been provided by
the Alfred P. Sloan Foundation, the U.S. Department of Energy Office of
Science, and the Participating Institutions. SDSS-IV acknowledges
support and resources from the Center for High-Performance Computing at
the University of Utah. The SDSS web site is www.sdss.org.

SDSS-IV is managed by the Astrophysical Research Consortium for the 
Participating Institutions of the SDSS Collaboration including the 
Brazilian Participation Group, the Carnegie Institution for Science, 
Carnegie Mellon University, the Chilean Participation Group, the French Participation Group, Harvard-Smithsonian Center for Astrophysics, 
Instituto de Astrof\'isica de Canarias, The Johns Hopkins University, 
Kavli Institute for the Physics and Mathematics of the Universe (IPMU) / 
University of Tokyo, Lawrence Berkeley National Laboratory, 
Leibniz Institut f\"ur Astrophysik Potsdam (AIP),  
Max-Planck-Institut f\"ur Astronomie (MPIA Heidelberg), 
Max-Planck-Institut f\"ur Astrophysik (MPA Garching), 
Max-Planck-Institut f\"ur Extraterrestrische Physik (MPE), 
National Astronomical Observatories of China, New Mexico State University, 
New York University, University of Notre Dame, 
Observat\'ario Nacional / MCTI, The Ohio State University, 
Pennsylvania State University, Shanghai Astronomical Observatory, 
United Kingdom Participation Group,
Universidad Nacional Aut\'onoma de M\'exico, University of Arizona, 
University of Colorado Boulder, University of Oxford, University of Portsmouth, 
University of Utah, University of Virginia, University of Washington, University of Wisconsin, 
Vanderbilt University, and Yale University.

We thank K. Verbeek for the use of the white dwarf-red dwarf model spectra which are partially made from white dwarf model spectra kindly given to us by D. Koester. 

J. van Roestel acknowledges support by the Netherlands Research School of Astronomy (NOVA) and Foundation for Fundamental Research on Matter (FOM). This paper was finalized during a stay by the lead authors at the Kavli Institute for Theoretical Physics, Santa Barbara, which is supported in part by the National Science Foundation under Grant No. NSF PHY-1125915.

\bibliographystyle{mnras}
\bibliography{PTFS1108ag} 

\begin{thebibliography}{}
\makeatletter
\relax
\def\mn@urlcharsother{\let\do\@makeother \do\$\do\&\do\#\do\^\do\_\do\%\do\~}
\def\mn@doi{\begingroup\mn@urlcharsother \@ifnextchar [ {\mn@doi@}
  {\mn@doi@[]}}
\def\mn@doi@[#1]#2{\def\@tempa{#1}\ifx\@tempa\@empty \href
  {http://dx.doi.org/#2} {doi:#2}\else \href {http://dx.doi.org/#2} {#1}\fi
  \endgroup}
\def\mn@eprint#1#2{\mn@eprint@#1:#2::\@nil}
\def\mn@eprint@arXiv#1{\href {http://arxiv.org/abs/#1} {{\tt arXiv:#1}}}
\def\mn@eprint@dblp#1{\href {http://dblp.uni-trier.de/rec/bibtex/#1.xml}
  {dblp:#1}}
\def\mn@eprint@#1:#2:#3:#4\@nil{\def\@tempa {#1}\def\@tempb {#2}\def\@tempc
  {#3}\ifx \@tempc \@empty \let \@tempc \@tempb \let \@tempb \@tempa \fi \ifx
  \@tempb \@empty \def\@tempb {arXiv}\fi \@ifundefined
  {mn@eprint@\@tempb}{\@tempb:\@tempc}{\expandafter \expandafter \csname
  mn@eprint@\@tempb\endcsname \expandafter{\@tempc}}}

\bibitem[\protect\citeauthoryear{{Ahn} et~al.,}{{Ahn}
  et~al.}{2012}]{2012ApJS..203...21A}
{Ahn} C.~P.,  et~al., 2012, \mn@doi [\apjs] {10.1088/0067-0049/203/2/21}, \href
  {http://adsabs.harvard.edu/abs/2012ApJS..203...21A} {203, 21}

\bibitem[\protect\citeauthoryear{{Baraffe}, {Homeier}, {Allard}  \&
  {Chabrier}}{{Baraffe} et~al.}{2015}]{2015A&A...577A..42B}
{Baraffe} I.,  {Homeier} D.,  {Allard} F.,   {Chabrier} G.,  2015, \mn@doi
  [\aap] {10.1051/0004-6361/201425481}, \href
  {http://adsabs.harvard.edu/abs/2015A%26A...577A..42B} {577, A42}

\bibitem[\protect\citeauthoryear{{Barman}, {Hauschildt}  \& {Allard}}{{Barman}
  et~al.}{2004}]{2004ApJ...614..338B}
{Barman} T.~S.,  {Hauschildt} P.~H.,   {Allard} F.,  2004, \mn@doi [\apj]
  {10.1086/423661}, \href {http://adsabs.harvard.edu/abs/2004ApJ...614..338B}
  {614, 338}

\bibitem[\protect\citeauthoryear{{Beatty} et~al.,}{{Beatty}
  et~al.}{2007}]{2007ApJ...663..573B}
{Beatty} T.~G.,  et~al., 2007, \mn@doi [\apj] {10.1086/518413}, \href
  {http://adsabs.harvard.edu/abs/2007ApJ...663..573B} {663, 573}

\bibitem[\protect\citeauthoryear{{Benvenuto} \& {Althaus}}{{Benvenuto} \&
  {Althaus}}{1999}]{1999MNRAS.303...30B}
{Benvenuto} O.~G.,  {Althaus} L.~G.,  1999, \mn@doi [\mnras]
  {10.1046/j.1365-8711.1999.02215.x}, \href
  {http://adsabs.harvard.edu/abs/1999MNRAS.303...30B} {303, 30}

\bibitem[\protect\citeauthoryear{{Bergeron} et~al.,}{{Bergeron}
  et~al.}{2011}]{2011ApJ...737...28B}
{Bergeron} P.,  et~al., 2011, \mn@doi [\apj] {10.1088/0004-637X/737/1/28},
  \href {http://adsabs.harvard.edu/abs/2011ApJ...737...28B} {737, 28}

\bibitem[\protect\citeauthoryear{{Beuermann}}{{Beuermann}}{2006}]{Beuermann:2006}
{Beuermann} K.,  2006, \mn@doi [\aap] {10.1051/0004-6361:20065930}, \href
  {http://adsabs.harvard.edu/abs/2006A%26A...460..783B} {460, 783}

\bibitem[\protect\citeauthoryear{{Beuermann}, {Dreizler}  \&
  {Hessman}}{{Beuermann} et~al.}{2013}]{2013A&A...555A.133B}
{Beuermann} K.,  {Dreizler} S.,   {Hessman} F.~V.,  2013, \mn@doi [\aap]
  {10.1051/0004-6361/201220510}, \href
  {http://adsabs.harvard.edu/abs/2013A%26A...555A.133B} {555, A133}

\bibitem[\protect\citeauthoryear{{Bours} et~al.,}{{Bours}
  et~al.}{2014}]{2014MNRAS.438.3399B}
{Bours} M.~C.~P.,  et~al., 2014, \mn@doi [\mnras] {10.1093/mnras/stt2453},
  \href {http://adsabs.harvard.edu/abs/2014MNRAS.438.3399B} {438, 3399}

\bibitem[\protect\citeauthoryear{{Chabrier}, {Gallardo}  \&
  {Baraffe}}{{Chabrier} et~al.}{2007}]{2007A&A...472L..17C}
{Chabrier} G.,  {Gallardo} J.,   {Baraffe} I.,  2007, \mn@doi [\aap]
  {10.1051/0004-6361:20077702}, \href
  {http://adsabs.harvard.edu/abs/2007A%26A...472L..17C} {472, L17}

\bibitem[\protect\citeauthoryear{{Claret} \& {Bloemen}}{{Claret} \&
  {Bloemen}}{2011}]{Claret:2011}
{Claret} A.,  {Bloemen} S.,  2011, \mn@doi [\aap]
  {10.1051/0004-6361/201116451}, \href
  {http://adsabs.harvard.edu/abs/2011A%26A...529A..75C} {529, A75}

\bibitem[\protect\citeauthoryear{{Copperwheat}, {Marsh}, {Dhillon},
  {Littlefair}, {Hickman}, {G{\"a}nsicke}  \& {Southworth}}{{Copperwheat}
  et~al.}{2010}]{2010MNRAS.402.1824C}
{Copperwheat} C.~M.,  {Marsh} T.~R.,  {Dhillon} V.~S.,  {Littlefair} S.~P.,
  {Hickman} R.,  {G{\"a}nsicke} B.~T.,   {Southworth} J.,  2010, \mn@doi
  [\mnras] {10.1111/j.1365-2966.2009.16010.x}, \href
  {http://adsabs.harvard.edu/abs/2010MNRAS.402.1824C} {402, 1824}

\bibitem[\protect\citeauthoryear{{Davis}, {Kolb}, {Willems}  \&
  {G{\"a}nsicke}}{{Davis} et~al.}{2008}]{2008MNRAS.389.1563D}
{Davis} P.~J.,  {Kolb} U.,  {Willems} B.,   {G{\"a}nsicke} B.~T.,  2008,
  \mn@doi [\mnras] {10.1111/j.1365-2966.2008.13675.x}, \href
  {http://adsabs.harvard.edu/abs/2008MNRAS.389.1563D} {389, 1563}

\bibitem[\protect\citeauthoryear{{Derekas} et~al.,}{{Derekas}
  et~al.}{2015}]{2015ApJ...808..179D}
{Derekas} A.,  et~al., 2015, \mn@doi [\apj] {10.1088/0004-637X/808/2/179},
  \href {http://adsabs.harvard.edu/abs/2015ApJ...808..179D} {808, 179}

\bibitem[\protect\citeauthoryear{{Dhillon} et~al.,}{{Dhillon}
  et~al.}{2007}]{2007MNRAS.378..825D}
{Dhillon} V.~S.,  et~al., 2007, \mn@doi [\mnras]
  {10.1111/j.1365-2966.2007.11881.x}, \href
  {http://adsabs.harvard.edu/abs/2007MNRAS.378..825D} {378, 825}

\bibitem[\protect\citeauthoryear{{Djorgovski} et~al.,}{{Djorgovski}
  et~al.}{2011}]{2011arXiv1102.5004D}
{Djorgovski} S.~G.,  et~al., 2011, preprint, \href
  {http://adsabs.harvard.edu/abs/2011arXiv1102.5004D} {} (\mn@eprint {arXiv}
  {1102.5004})

\bibitem[\protect\citeauthoryear{{Drake} et~al.,}{{Drake}
  et~al.}{2009}]{2009ApJ...696..870D}
{Drake} A.~J.,  et~al., 2009, \mn@doi [\apj] {10.1088/0004-637X/696/1/870},
  \href {http://adsabs.harvard.edu/abs/2009ApJ...696..870D} {696, 870}

\bibitem[\protect\citeauthoryear{{Eastman}, {Siverd}  \& {Gaudi}}{{Eastman}
  et~al.}{2010}]{2010PASP..122..935E}
{Eastman} J.,  {Siverd} R.,   {Gaudi} B.~S.,  2010, \mn@doi [\pasp]
  {10.1086/655938}, \href {http://adsabs.harvard.edu/abs/2010PASP..122..935E}
  {122, 935}

\bibitem[\protect\citeauthoryear{{Feline}, {Dhillon}, {Marsh}  \&
  {Brinkworth}}{{Feline} et~al.}{2004}]{2004MNRAS.355....1F}
{Feline} W.~J.,  {Dhillon} V.~S.,  {Marsh} T.~R.,   {Brinkworth} C.~S.,  2004,
  \mn@doi [\mnras] {10.1111/j.1365-2966.2004.08302.x}, \href
  {http://adsabs.harvard.edu/abs/2004MNRAS.355....1F} {355, 1}

\bibitem[\protect\citeauthoryear{{Fontaine}, {Brassard}  \&
  {Bergeron}}{{Fontaine} et~al.}{2001}]{2001PASP..113..409F}
{Fontaine} G.,  {Brassard} P.,   {Bergeron} P.,  2001, \mn@doi [\pasp]
  {10.1086/319535}, \href {http://adsabs.harvard.edu/abs/2001PASP..113..409F}
  {113, 409}

\bibitem[\protect\citeauthoryear{{Foreman-Mackey}, {Hogg}, {Lang}  \&
  {Goodman}}{{Foreman-Mackey} et~al.}{2013}]{2013PASP..125..306F}
{Foreman-Mackey} D.,  {Hogg} D.~W.,  {Lang} D.,   {Goodman} J.,  2013, \mn@doi
  [\pasp] {10.1086/670067}, \href
  {http://adsabs.harvard.edu/abs/2013PASP..125..306F} {125, 306}

\bibitem[\protect\citeauthoryear{{G{\"a}nsicke} et~al.,}{{G{\"a}nsicke}
  et~al.}{2009}]{2009MNRAS.397.2170G}
{G{\"a}nsicke} B.~T.,  et~al., 2009, \mn@doi [\mnras]
  {10.1111/j.1365-2966.2009.15126.x}, \href
  {http://adsabs.harvard.edu/abs/2009MNRAS.397.2170G} {397, 2170}

\bibitem[\protect\citeauthoryear{{Gianninas}, {Strickland}, {Kilic}  \&
  {Bergeron}}{{Gianninas} et~al.}{2013}]{2013ApJ...766....3G}
{Gianninas} A.,  {Strickland} B.~D.,  {Kilic} M.,   {Bergeron} P.,  2013,
  \mn@doi [\apj] {10.1088/0004-637X/766/1/3}, \href
  {http://adsabs.harvard.edu/abs/2013ApJ...766....3G} {766, 3}

\bibitem[\protect\citeauthoryear{{Horne} \& {Schneider}}{{Horne} \&
  {Schneider}}{1989}]{1989ApJ...343..888H}
{Horne} K.,  {Schneider} D.~P.,  1989, \mn@doi [\apj] {10.1086/167758}, \href
  {http://adsabs.harvard.edu/abs/1989ApJ...343..888H} {343, 888}

\bibitem[\protect\citeauthoryear{{Howell}, {Nelson}  \& {Rappaport}}{{Howell}
  et~al.}{2001}]{2001ApJ...550..897H}
{Howell} S.~B.,  {Nelson} L.~A.,   {Rappaport} S.,  2001, \mn@doi [\apj]
  {10.1086/319776}, \href {http://adsabs.harvard.edu/abs/2001ApJ...550..897H}
  {550, 897}

\bibitem[\protect\citeauthoryear{{Ivanova} et~al.,}{{Ivanova}
  et~al.}{2013}]{2013A&ARv..21...59I}
{Ivanova} N.,  et~al., 2013, \mn@doi [\aapr] {10.1007/s00159-013-0059-2}, \href
  {http://adsabs.harvard.edu/abs/2013A%26ARv..21...59I} {21, 59}

\bibitem[\protect\citeauthoryear{{Knigge}}{{Knigge}}{2006}]{2006MNRAS.373..484K}
{Knigge} C.,  2006, \mn@doi [\mnras] {10.1111/j.1365-2966.2006.11096.x}, \href
  {http://adsabs.harvard.edu/abs/2006MNRAS.373..484K} {373, 484}

\bibitem[\protect\citeauthoryear{Knigge, Baraffe  \& Patterson}{Knigge
  et~al.}{2011}]{Knigge:2011}
Knigge C.,  Baraffe I.,   Patterson J.,  2011, \mn@doi [Astrophys.J.Suppl.]
  {10.1088/0067-0049/194/2/28}, 194, 28

\bibitem[\protect\citeauthoryear{{Koester} et~al.,}{{Koester}
  et~al.}{2001}]{koester:2001}
{Koester} D.,  et~al., 2001, \mn@doi [\aap] {10.1051/0004-6361:20011235}, \href
  {http://adsabs.harvard.edu/abs/2001A%26A...378..556K} {378, 556}

\bibitem[\protect\citeauthoryear{{Kupfer} et~al.,}{{Kupfer}
  et~al.}{2015}]{2015A&A...576A..44K}
{Kupfer} T.,  et~al., 2015, \mn@doi [\aap] {10.1051/0004-6361/201425213}, \href
  {http://adsabs.harvard.edu/abs/2015A%26A...576A..44K} {576, A44}

\bibitem[\protect\citeauthoryear{{Laher} et~al.,}{{Laher}
  et~al.}{2014}]{2014PASP..126..674L}
{Laher} R.~R.,  et~al., 2014, \mn@doi [\pasp] {10.1086/677351}, \href
  {http://adsabs.harvard.edu/abs/2014PASP..126..674L} {126, 674}

\bibitem[\protect\citeauthoryear{{Law} et~al.,}{{Law} et~al.}{2009}]{Law:2009}
{Law} N.~M.,  et~al., 2009, \mn@doi [\pasp] {10.1086/648598}, \href
  {http://adsabs.harvard.edu/abs/2009PASP..121.1395L} {121, 1395}

\bibitem[\protect\citeauthoryear{{L{\'o}pez-Morales}}{{L{\'o}pez-Morales}}{2007}]{2007ApJ...660..732L}
{L{\'o}pez-Morales} M.,  2007, \mn@doi [\apj] {10.1086/513142}, \href
  {http://adsabs.harvard.edu/abs/2007ApJ...660..732L} {660, 732}

\bibitem[\protect\citeauthoryear{{Marsh} et~al.,}{{Marsh}
  et~al.}{2014}]{2014MNRAS.437..475M}
{Marsh} T.~R.,  et~al., 2014, \mn@doi [\mnras] {10.1093/mnras/stt1903}, \href
  {http://adsabs.harvard.edu/abs/2014MNRAS.437..475M} {437, 475}

\bibitem[\protect\citeauthoryear{{Moffat}}{{Moffat}}{1969}]{1969A&A.....3..455M}
{Moffat} A.~F.~J.,  1969, \aap, \href
  {http://adsabs.harvard.edu/abs/1969A%26A.....3..455M} {3, 455}

\bibitem[\protect\citeauthoryear{{Nefs} et~al.,}{{Nefs}
  et~al.}{2013}]{2013MNRAS.431.3240N}
{Nefs} S.~V.,  et~al., 2013, \mn@doi [\mnras] {10.1093/mnras/stt405}, \href
  {http://adsabs.harvard.edu/abs/2013MNRAS.431.3240N} {431, 3240}

\bibitem[\protect\citeauthoryear{{Oke} \& {Gunn}}{{Oke} \&
  {Gunn}}{1982}]{1982PASP...94..586O}
{Oke} J.~B.,  {Gunn} J.~E.,  1982, \mn@doi [\pasp] {10.1086/131027}, \href
  {http://adsabs.harvard.edu/abs/1982PASP...94..586O} {94, 586}

\bibitem[\protect\citeauthoryear{{Paczynski}}{{Paczynski}}{1976}]{1976IAUS...73...75P}
{Paczynski} B.,  1976, in {Eggleton} P.,  {Mitton} S.,   {Whelan} J.,  eds,
  IAU Symposium Vol. 73, Structure and Evolution of Close Binary Systems. p.~75

\bibitem[\protect\citeauthoryear{{Pala} et~al.,}{{Pala}
  et~al.}{2017}]{2017MNRAS.466.2855P}
{Pala} A.~F.,  et~al., 2017, \mn@doi [\mnras] {10.1093/mnras/stw3293}, \href
  {http://adsabs.harvard.edu/abs/2017MNRAS.466.2855P} {466, 2855}

\bibitem[\protect\citeauthoryear{{Panei}, {Althaus}, {Chen}  \& {Han}}{{Panei}
  et~al.}{2007}]{2007MNRAS.382..779P}
{Panei} J.~A.,  {Althaus} L.~G.,  {Chen} X.,   {Han} Z.,  2007, \mn@doi
  [\mnras] {10.1111/j.1365-2966.2007.12400.x}, \href
  {http://adsabs.harvard.edu/abs/2007MNRAS.382..779P} {382, 779}

\bibitem[\protect\citeauthoryear{{Parsons}, {Marsh}, {Copperwheat}, {Dhillon},
  {Littlefair}, {G{\"a}nsicke}  \& {Hickman}}{{Parsons}
  et~al.}{2010a}]{Parsons:2009}
{Parsons} S.~G.,  {Marsh} T.~R.,  {Copperwheat} C.~M.,  {Dhillon} V.~S.,
  {Littlefair} S.~P.,  {G{\"a}nsicke} B.~T.,   {Hickman} R.,  2010a, \mn@doi
  [\mnras] {10.1111/j.1365-2966.2009.16072.x}, \href
  {http://adsabs.harvard.edu/abs/2010MNRAS.402.2591P} {402, 2591}

\bibitem[\protect\citeauthoryear{{Parsons} et~al.,}{{Parsons}
  et~al.}{2010b}]{2010MNRAS.407.2362P}
{Parsons} S.~G.,  et~al., 2010b, \mn@doi [\mnras]
  {10.1111/j.1365-2966.2010.17063.x}, \href
  {http://adsabs.harvard.edu/abs/2010MNRAS.407.2362P} {407, 2362}

\bibitem[\protect\citeauthoryear{{Parsons} et~al.,}{{Parsons}
  et~al.}{2012a}]{Parsons:2011a}
{Parsons} S.~G.,  et~al., 2012a, \mn@doi [\mnras]
  {10.1111/j.1365-2966.2011.19691.x}, \href
  {http://adsabs.harvard.edu/abs/2012MNRAS.419..304P} {419, 304}

\bibitem[\protect\citeauthoryear{{Parsons} et~al.,}{{Parsons}
  et~al.}{2012b}]{2012MNRAS.420.3281P}
{Parsons} S.~G.,  et~al., 2012b, \mn@doi [\mnras]
  {10.1111/j.1365-2966.2011.20251.x}, \href
  {http://adsabs.harvard.edu/abs/2012MNRAS.420.3281P} {420, 3281}

\bibitem[\protect\citeauthoryear{{Parsons} et~al.,}{{Parsons}
  et~al.}{2012c}]{2012MNRAS.426.1950P}
{Parsons} S.~G.,  et~al., 2012c, \mn@doi [\mnras]
  {10.1111/j.1365-2966.2012.21773.x}, \href
  {http://adsabs.harvard.edu/abs/2012MNRAS.426.1950P} {426, 1950}

\bibitem[\protect\citeauthoryear{{Parsons} et~al.,}{{Parsons}
  et~al.}{2015}]{2015MNRAS.449.2194P}
{Parsons} S.~G.,  et~al., 2015, \mn@doi [\mnras] {10.1093/mnras/stv382}, \href
  {http://adsabs.harvard.edu/abs/2015MNRAS.449.2194P} {449, 2194}

\bibitem[\protect\citeauthoryear{{Parsons} et~al.,}{{Parsons}
  et~al.}{2016}]{2016MNRAS.458.2793P}
{Parsons} S.~G.,  et~al., 2016, \mn@doi [\mnras] {10.1093/mnras/stw516}, \href
  {http://adsabs.harvard.edu/abs/2016MNRAS.458.2793P} {458, 2793}

\bibitem[\protect\citeauthoryear{{Peters}}{{Peters}}{1964}]{1964PhRv..136.1224P}
{Peters} P.~C.,  1964, \mn@doi [Physical Review] {10.1103/PhysRev.136.B1224},
  \href {http://adsabs.harvard.edu/abs/1964PhRv..136.1224P} {136, 1224}

\bibitem[\protect\citeauthoryear{{Peters} \& {Mathews}}{{Peters} \&
  {Mathews}}{1963}]{1963PhRv..131..435P}
{Peters} P.~C.,  {Mathews} J.,  1963, \mn@doi [Physical Review]
  {10.1103/PhysRev.131.435}, \href
  {http://adsabs.harvard.edu/abs/1963PhRv..131..435P} {131, 435}

\bibitem[\protect\citeauthoryear{{Pickles}}{{Pickles}}{1998}]{Pickles:1998}
{Pickles} A.~J.,  1998, \mn@doi [\pasp] {10.1086/316197}, \href
  {http://adsabs.harvard.edu/abs/1998PASP..110..863P} {110, 863}

\bibitem[\protect\citeauthoryear{{Pyrzas} et~al.,}{{Pyrzas}
  et~al.}{2012}]{Pyrzas:2011bk}
{Pyrzas} S.,  et~al., 2012, \mn@doi [\mnras]
  {10.1111/j.1365-2966.2011.19746.x}, \href
  {http://adsabs.harvard.edu/abs/2012MNRAS.419..817P} {419, 817}

\bibitem[\protect\citeauthoryear{{Rajpurohit}, {Reyl{\'e}}, {Allard},
  {Homeier}, {Schultheis}, {Bessell}  \& {Robin}}{{Rajpurohit}
  et~al.}{2013}]{2013A&A...556A..15R}
{Rajpurohit} A.~S.,  {Reyl{\'e}} C.,  {Allard} F.,  {Homeier} D.,  {Schultheis}
  M.,  {Bessell} M.~S.,   {Robin} A.~C.,  2013, \mn@doi [\aap]
  {10.1051/0004-6361/201321346}, \href
  {http://adsabs.harvard.edu/abs/2013A%26A...556A..15R} {556, A15}

\bibitem[\protect\citeauthoryear{{Rappaport}, {Verbunt}  \& {Joss}}{{Rappaport}
  et~al.}{1983}]{1983ApJ...275..713R}
{Rappaport} S.,  {Verbunt} F.,   {Joss} P.~C.,  1983, \mn@doi [\apj]
  {10.1086/161569}, \href {http://adsabs.harvard.edu/abs/1983ApJ...275..713R}
  {275, 713}

\bibitem[\protect\citeauthoryear{{Rau} et~al.,}{{Rau} et~al.}{2009}]{Rau:2009}
{Rau} A.,  et~al., 2009, \mn@doi [\pasp] {10.1086/605911}, \href
  {http://adsabs.harvard.edu/abs/2009PASP..121.1334R} {121, 1334}

\bibitem[\protect\citeauthoryear{{Rebassa-Mansergas}, {G{\"a}nsicke},
  {Rodr{\'{\i}}guez-Gil}, {Schreiber}  \& {Koester}}{{Rebassa-Mansergas}
  et~al.}{2007}]{2007MNRAS.382.1377R}
{Rebassa-Mansergas} A.,  {G{\"a}nsicke} B.~T.,  {Rodr{\'{\i}}guez-Gil} P.,
  {Schreiber} M.~R.,   {Koester} D.,  2007, \mn@doi [\mnras]
  {10.1111/j.1365-2966.2007.12288.x}, \href
  {http://adsabs.harvard.edu/abs/2007MNRAS.382.1377R} {382, 1377}

\bibitem[\protect\citeauthoryear{{Ribas}}{{Ribas}}{2006}]{2006Ap&SS.304...89R}
{Ribas} I.,  2006, \mn@doi [\apss] {10.1007/s10509-006-9081-4}, \href
  {http://adsabs.harvard.edu/abs/2006Ap%26SS.304...89R} {304, 89}

\bibitem[\protect\citeauthoryear{{Ritter} \& {Kolb}}{{Ritter} \&
  {Kolb}}{2003}]{2003A&A...404..301R}
{Ritter} H.,  {Kolb} U.,  2003, \mn@doi [\aap] {10.1051/0004-6361:20030330},
  \href {http://adsabs.harvard.edu/abs/2003A%26A...404..301R} {404, 301}

\bibitem[\protect\citeauthoryear{{Ritter}, {Zhang}  \& {Kolb}}{{Ritter}
  et~al.}{2000}]{2000A&A...360..969R}
{Ritter} H.,  {Zhang} Z.-Y.,   {Kolb} U.,  2000, \aap, \href
  {http://adsabs.harvard.edu/abs/2000A%26A...360..969R} {360, 959}

\bibitem[\protect\citeauthoryear{{Schlegel}, {Finkbeiner}  \&
  {Davis}}{{Schlegel} et~al.}{1998}]{1998ApJ...500..525S}
{Schlegel} D.~J.,  {Finkbeiner} D.~P.,   {Davis} M.,  1998, \mn@doi [\apj]
  {10.1086/305772}, \href {http://adsabs.harvard.edu/abs/1998ApJ...500..525S}
  {500, 525}

\bibitem[\protect\citeauthoryear{{Schreiber} et~al.,}{{Schreiber}
  et~al.}{2010}]{2010A&A...513L...7S}
{Schreiber} M.~R.,  et~al., 2010, \mn@doi [\aap] {10.1051/0004-6361/201013990},
  \href {http://adsabs.harvard.edu/abs/2010A%26A...513L...7S} {513, L7}

\bibitem[\protect\citeauthoryear{{Shears}, {Brady}, {Gaensicke}, {Krajci},
  {Miller}, {Oegmen}, {Pietz}  \& {Staels}}{{Shears}
  et~al.}{2009}]{2009JBAA..119..144S}
{Shears} J.,  {Brady} S.,  {Gaensicke} B.,  {Krajci} T.,  {Miller} I.,
  {Oegmen} Y.,  {Pietz} J.,   {Staels} B.,  2009, Journal of the British
  Astronomical Association, \href
  {http://adsabs.harvard.edu/abs/2009JBAA..119..144S} {119, 144}

\bibitem[\protect\citeauthoryear{{Stehle}, {Ritter}  \& {Kolb}}{{Stehle}
  et~al.}{1996}]{1996MNRAS.279..581S}
{Stehle} R.,  {Ritter} H.,   {Kolb} U.,  1996, \mn@doi [\mnras]
  {10.1093/mnras/279.2.581}, \href
  {http://adsabs.harvard.edu/abs/1996MNRAS.279..581S} {279, 581}

\bibitem[\protect\citeauthoryear{{Taam} \& {Ricker}}{{Taam} \&
  {Ricker}}{2010}]{2010NewAR..54...65T}
{Taam} R.~E.,  {Ricker} P.~M.,  2010, \mn@doi [\nar]
  {10.1016/j.newar.2010.09.027}, \href
  {http://adsabs.harvard.edu/abs/2010NewAR..54...65T} {54, 65}

\bibitem[\protect\citeauthoryear{{Taam} \& {Sandquist}}{{Taam} \&
  {Sandquist}}{2000}]{2000ARA&A..38..113T}
{Taam} R.~E.,  {Sandquist} E.~L.,  2000, \mn@doi [\araa]
  {10.1146/annurev.astro.38.1.113}, \href
  {http://adsabs.harvard.edu/abs/2000ARA%26A..38..113T} {38, 113}

\bibitem[\protect\citeauthoryear{{Tal-Or} et~al.,}{{Tal-Or}
  et~al.}{2013}]{2013A&A...553A..30T}
{Tal-Or} L.,  et~al., 2013, \mn@doi [\aap] {10.1051/0004-6361/201220862}, \href
  {http://adsabs.harvard.edu/abs/2013A%26A...553A..30T} {553, A30}

\bibitem[\protect\citeauthoryear{{Townsley} \& {G{\"a}nsicke}}{{Townsley} \&
  {G{\"a}nsicke}}{2009}]{2009ApJ...693.1007T}
{Townsley} D.~M.,  {G{\"a}nsicke} B.~T.,  2009, \mn@doi [\apj]
  {10.1088/0004-637X/693/1/1007}, \href
  {http://adsabs.harvard.edu/abs/2009ApJ...693.1007T} {693, 1007}

\bibitem[\protect\citeauthoryear{{VanderPlas} \& {Ivezic}}{{VanderPlas} \&
  {Ivezic}}{2015}]{VanDerPlas:2015}
{VanderPlas} J.~T.,  {Ivezic} Z.,  2015, preprint, \href
  {http://adsabs.harvard.edu/abs/2015arXiv150201344V} {} (\mn@eprint {arXiv}
  {1502.01344})

\bibitem[\protect\citeauthoryear{{Verbeek} et~al.,}{{Verbeek}
  et~al.}{2014}]{verbeek:2014}
{Verbeek} K.,  et~al., 2014, \mn@doi [\mnras] {10.1093/mnras/stt1492}, \href
  {http://adsabs.harvard.edu/abs/2014MNRAS.438....2V} {438, 2}

\bibitem[\protect\citeauthoryear{{Verbunt} \& {Rappaport}}{{Verbunt} \&
  {Rappaport}}{1988}]{1988ApJ...332..193V}
{Verbunt} F.,  {Rappaport} S.,  1988, \mn@doi [\apj] {10.1086/166645}, \href
  {http://adsabs.harvard.edu/abs/1988ApJ...332..193V} {332, 193}

\bibitem[\protect\citeauthoryear{{Wade} \& {Horne}}{{Wade} \&
  {Horne}}{1988}]{Wade:1988}
{Wade} R.~A.,  {Horne} K.,  1988, \mn@doi [\apj] {10.1086/165905}, \href
  {http://adsabs.harvard.edu/abs/1988ApJ...324..411W} {324, 411}

\bibitem[\protect\citeauthoryear{{Wall} \& {Jenkins}}{{Wall} \&
  {Jenkins}}{2012}]{2012psa..book.....W}
{Wall} J.~V.,  {Jenkins} C.~R.,  2012, {Practical Statistics for Astronomers}

\bibitem[\protect\citeauthoryear{{Webbink}}{{Webbink}}{2008}]{2008ASSL..352..233W}
{Webbink} R.~F.,  2008, in {Milone} E.~F.,  {Leahy} D.~A.,   {Hobill} D.~W.,
  eds,  Astrophysics and Space Science Library Vol. 352, Astrophysics and Space
  Science Library. p.~233 (\mn@eprint {arXiv} {0704.0280})

\bibitem[\protect\citeauthoryear{{Wood}}{{Wood}}{1995}]{Wood:1995}
{Wood} M.~A.,  1995, in {Koester} D.,  {Werner} K.,  eds,  Lecture Notes in
  Physics, Berlin Springer Verlag Vol. 443, White Dwarfs. p.~41,
  \mn@doi{10.1007/3-540-59157-5_171}

\bibitem[\protect\citeauthoryear{{Zhou} et~al.,}{{Zhou}
  et~al.}{2014}]{2014MNRAS.437.2831Z}
{Zhou} G.,  et~al., 2014, \mn@doi [\mnras] {10.1093/mnras/stt2100}, \href
  {http://adsabs.harvard.edu/abs/2014MNRAS.437.2831Z} {437, 2831}

\bibitem[\protect\citeauthoryear{{Zorotovic} et~al.,}{{Zorotovic}
  et~al.}{2016}]{2016arXiv160107785Z}
{Zorotovic} M.,  et~al., 2016, \mn@doi [\mnras] {10.1093/mnras/stw246}, \href
  {http://adsabs.harvard.edu/abs/2016MNRAS.457.3867Z} {457, 3867}

\bibitem[\protect\citeauthoryear{van Dokkum}{van
  Dokkum}{2001}]{vanDokkum:2001bp}
van Dokkum P.~G.,  2001, \mn@doi [Publ.Astron.Soc.Pac.] {10.1086/323894}, 113,
  1420

\makeatother
\end{thebibliography}

\appendix

\section{Centre of light offset}
\label{appendix}
For stars that do not have a uniform surface brightness, the centre of light emitted by that star is not the same as the centre of mass of the star. If the star is in a close binary, this can significantly offset the measured radial velocity amplitude. We describe this effect using $f\equiv \mathrm{offset}/R$, the distance between the centre of light from the centre of mass divided by the radius of the star. If $f=0$, the offset is zero, while $f=1$ is the most extreme case, where the centre of light is at the surface of the star. In this Appendix we discuss the value of $f$ at quadrature, starting from a simple model.

\citet{Wade:1988} discussed the case where one hemisphere of the star is showing an absorption line. In this case, assuming the star is spherical, $f$ is equivalent to the centre of mass of a semi-circle:
\begin{equation}\label{eq:semicircle}
f=\dfrac{\int_{0}^{\pi/2} \cos(\theta) \sin^2(\theta) d\theta }{\int_{0}^{\pi/2}  \sin^2(\theta) d\theta } = \dfrac{4}{3\pi}\approx0.424
\end{equation}

In the paper by \citet{1989ApJ...343..888H}, a lower limit is calculated for the offset factor of an irradiated red dwarf. Assuming a spherical star, an irradiation source at infinity, and the reflected light is proportional to the incoming light, the value of $f$ can be calculated by analytically solving equation \ref{eq:sphere_infinitedistance}.
\begin{equation}\label{eq:sphere_infinitedistance}
f=\dfrac{\int_{0}^{\pi/2} \cos^2(\theta) \sin^2(\theta) d\theta }{\int_{0}^{\pi/2} \cos(\theta) \sin^2(\theta) d\theta } = \dfrac{3\pi}{16} \approx 0.589
\end{equation}

If the reflection effect is dominant, this usually means that the irradiated star's radius is significant compared to the semi-major axis, and the assumption that the irradiation source is at infinity is not correct any more. Assuming a point source at a finite distance has two effects; the irradiation is stronger closer to the irradiating object, and not the entire hemisphere is irradiated. Equation \ref{eq:sphere_finitedistance} includes these two corrections, and Fig.~\ref{fig:schematic} shows what all components mean. The solution of the equation is only a function of $r$, which is shown in Fig.~\ref{fig:correction_effect} by the black line.

\def\centerarc[#1](#2)(#3:#4:#5)
	{ \draw[#1] ($(#2)+({#5*cos(#3)},{#5*sin(#3)})$) arc (#3:#4:#5); }
\newcommand{\AxisRotator}[1][rotate=0]{%
    \tikz [x=0.25cm,y=0.60cm,line width=.2ex,-stealth,#1] \draw (0,0) arc (-150:150:1 and 1);%
}

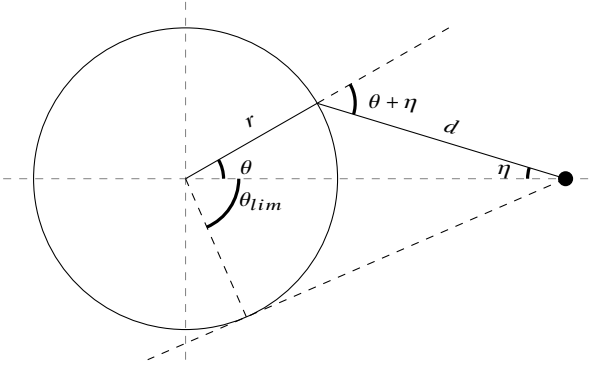
\begin{figure}
\begin{tikzpicture}
\newcommand*{\x}{30.}%
\newcommand*{\y}{17.0142316}%
\newcommand*{\limangle}{66.42}%

 \draw[gray,very thin,dashed] (-2.4,0) -- (5.4,0);
 \draw[gray,very thin,dashed] (0,-2.4) -- (0,2.4);

 \draw (0,0) circle (2cm);

 \draw (5,0) node[circle,fill,inner sep=2pt]{};

 \draw [color=black](0.87,0.77) node[rotate=\x] {$r$};
 \draw [color=black](3.5,0.7) node[rotate=-\y] {$d$};
 
 \draw [color=black](0.8,0.15) node[rotate=0] {$\theta$};
 \centerarc[black,very thick](0,0)(0:\x:0.5)
 
 \draw [color=black](4.2,0.1) node[rotate=0] {$\eta$};
 \centerarc[black,very thick](5,0)(180:180-\y:0.5)
 
 \draw [color=black](1.7320,1.0)+(1.0,0) node[rotate=0] {$\theta+\eta$};
 \centerarc[black,very thick](1.7320,1.0)(-\y:\x:0.5)

 \draw (0,0) -- (1.7320,1.0);
 \draw (5,0) -- (1.7320,1.0);
 \draw [dashed](1.7320,1.0) -- (1.7320*2,1.0*2);
 
 
 \draw [color=black](1.0,-0.3) node[rotate=0] {$\theta_{lim}$};
 \centerarc[black,very thick](0.,0.)(0:-\limangle:0.7)
 
 \draw [dashed] (5,0) -- ($(5,0)+({-6*sin(\limangle)},{-6*cos(\limangle)})$);
  \draw [dashed] (0,0) -- ($(0,0)+({2*cos(\limangle)},{-2*sin(\limangle)})$);

  \end{tikzpicture}
  \caption{This figure shows the angles involved in the calculation. The circle represents the irradiated star's surface and the dot indicates the point source.}
 \label{fig:schematic}
 \end{figure}

 \begin{equation}\label{eq:sphere_finitedistance}
f=\dfrac{\int_{0}^{\theta_{\rm lim}} \cos(\theta) d^{-2} \cos(\theta+\eta)  \sin^2(\theta) d\theta }{\int_{0}^{\theta_{\rm lim}} d^{-2} \cos(\theta+\eta) \sin^2(\theta) d\theta}
\end{equation}

\begin{align*}
  \text{with}~\eta &= \arctan \dfrac{r \sin\theta}{1-r\cos\theta}, \\
  d &= \dfrac{r\sin\theta }{ \sin\eta}, \text{ and } \\
 \theta_{\rm lim} &= \pi/2-\arcsin(r)
\end{align*}

So far we have assumed a spherical star, which makes it easy to write down a numerical expression for $f$. However, stars in close binaries are deformed by their companion star into a Roche geometry. This makes the analytic solution extremely complicated and solving the problem requires numerically integrating the incoming light and the observing angle over the entire surface of the star. The most extreme deviation is for Roche lobe filling stars, shown by the grey line in Fig.~\ref{fig:correction_effect}. As you can see, taking this into account decreases the offset factor compared to a spherical star. The reasons is that due to the Roche geometry, the maximum radius increases, while the rest of the star remains approximately spherical, and thus the factor $f$ decreases.

\begin{figure}
\begin{center}
\includegraphics[width=84mm]{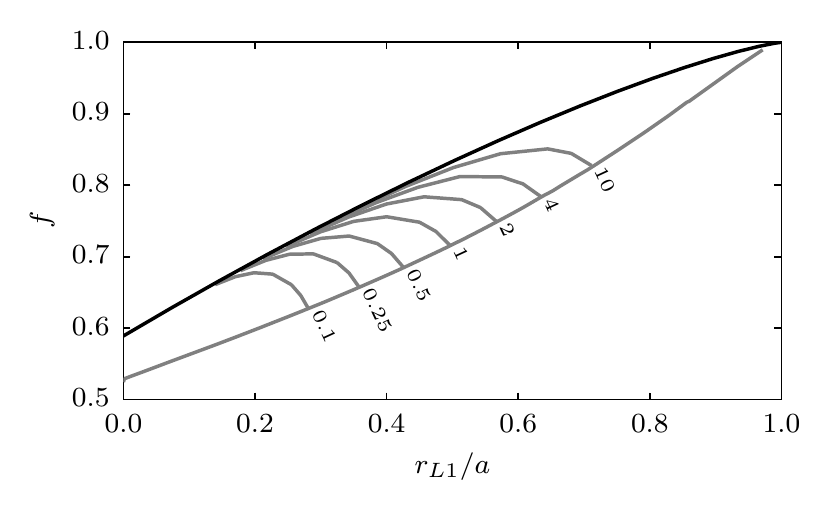}
\caption{The offset factor $f$ versus the radius of the star divided by the orbital separation. The black line indicates an irradiated sphere by a point source at a finite distance. The grey line indicates the offset factor for irradiated stars filling their Roche lobe. The number next to the grey line indicate the mass ratio $q$.}
\label{fig:correction_effect}
\end{center}
\end{figure}

\bsp	
\label{lastpage}
\end{document}